\documentclass[aps,prd,twocolumn,showpacs,10pt,superscriptaddress,nofootinbib,preprintnumbers,showkeys]{revtex4-1}
\usepackage{epsfig,amssymb,amsmath,psfrag,epstopdf,color}
\pdfoutput=1
\usepackage{graphicx}
\usepackage{multirow}
\usepackage{ulem}
\usepackage[breaklinks,bookmarksopen=true]{hyperref}
\usepackage{natbib}
\definecolor{linkcolor}{rgb}{0.0,0.3,0.5}
\usepackage[all]{hypcap}
\usepackage[T1]{fontenc}
\usepackage[utf8]{inputenc}
\usepackage{slashed}
\usepackage{booktabs}
\usepackage{lipsum}
\usepackage[table]{xcolor}
\usepackage{comment}
\usepackage{adjustbox,array}
\usepackage[section]{placeins} 
\usepackage{hyperref}

\definecolor{rosso}{cmyk}{0,1,1,0.4}
\definecolor{rossos}{cmyk}{0,1,1,0.55}
\definecolor{rossoc}{cmyk}{0,1,1,0.2}
\definecolor{blu}{cmyk}{1,1,0,0.3}
\definecolor{blus}{cmyk}{1,1,0,0.6}
\definecolor{bluc}{cmyk}{1,1,0,0.1}
\definecolor{verde}{cmyk}{0.92,0,0.59,0.25}
\definecolor{verdec}{cmyk}{0.92,0,0.59,0.15}
\definecolor{verdes}{cmyk}{0.92,0,0.59,0.4}

\DeclareUnicodeCharacter{2212}{\ensuremath{-}}

\hypersetup{
 colorlinks=true,
 linkcolor=verdec,
 urlcolor=verdec,
 citecolor=verdec
}

\begin{document}

\preprint{CTPU-PTC-2023-02}

\title{Lepton portal dark matter at muon colliders:  Total rates and generic features for phenomenologically viable scenarios}

\def\andname{\hspace*{-0.5em}} 
\author{Adil Jueid}
\email{adiljueid@ibs.re.kr}
\affiliation{Center for Theoretical Physics of the Universe, Institute for Basic Science (IBS), 34126 Daejeon, Republic of Korea}
\author{and Salah Nasri}
\email{snasri@uaeu.ac.ae; salah.nasri@cern.ch}
\affiliation{Department of physics, United Arab Emirates University, Al-Ain, UAE}
\affiliation{The Abdus Salam International Centre for Theoretical Physics, Strada Costiera 11, I-34014, Trieste, Italy}

\begin{abstract}
Lepton portal dark matter (DM) models are a class of models where the DM candidates solely couple to charged leptons through a mediator carrying a lepton number. These models are very interesting since they avoid constraints from direct detection experiments even for coupling of order ${\cal O}(1)$, they have small annihilation cross sections,  and can be probed efficiently at lepton colliders.  In this work, we consider a minimal lepton portal DM model which consists of extending the SM with two $SU(2)_L$ singlets: a charged scalar singlet and an electrically  neutral right-handed fermion. We systematically study the production mechanisms of DM at multi-TeV muon colliders. After considering  all the possible theoretical and experimental constraints and studying the phenomenology of lepton flavour violation and DM in the muon-philic scenario, we analyse the production rates of 54 channels (26 channels for prompt DM production and 28 channels for charged scalar production) at multi-TeV muon colliders. Finally, we discuss the possible collider signatures of some channels and the corresponding backgrounds. We find  that at least 9 channels for DM production can be very efficient in testing DM with masses up to about $1$ TeV.
\end{abstract}

\maketitle


\newcommand{\ssection}[1]{{\em #1.\ }}
\newcommand{\rem}[1]{\textbf{#1}}

\section{Introduction}
\label{sec:Intro}

Supported by various astrophysical and cosmological observations, it is now widely accepted that dark matter (DM) exists in the universe (see {\it e.g.} \cite{Jungman:1995df,Bergstrom:2000pn,Bertone:2004pz,Feng:2010gw} for comprehensive reviews). On the other hand, the measurements of the anisotropies in the cosmic microwave background (CMB) implies that DM is the dominant component of the matter budget in the universe with a density of $\Omega_{\rm DM} h^2 = 0.1198\pm0.0015$ \cite{Ade:2015xua}. The standard theories of structure formation require that the DM should be non-relativistic at the matter-radiation equality. In particle physics models, this can be easily realised by extending the SM with weakly interacting massive particles (WIMPs) under the standard thermal freeze-out mechanism.  The search for WIMPs  was one of the major programmes at the Large Hadron Collider (LHC). A special characteristic of WIMPs production at the LHC is that one can probe it through the recoil of a SM particle against a large missing transverse energy ($E_T^{\rm miss}$). Examples of these processes are mono-jet \cite{Beltran:2010ww}, mono-$Z$ \cite{Carpenter:2012rg,Bell:2012rg} or mono-Higgs \cite{Berlin:2014cfa} among others. Unfortunately various  searches for WIMPs at the LHC were unsuccessful to find such signals and limits were put on the production cross section versus the DM mass \cite{ATLAS:2021kxv,ATLAS:2021shl,ATLAS:2021hza,CMS:2019ykj,CMS:2020ulv} which were interpreted in various particle physics realizations. Furthermore, these constraints were even more stringent when the void bounds from direct detection experiments \cite{Aprile:2018dbl, Cui:2017nnn} are included \cite{GAMBIT:2017gge,GAMBIT:2017zdo}. The situation is not very different in the case the DM production is mediated through colored mediators or leptoquarks with the main mechanisms for DM density in the early universe being the co-annihilation or conversion-driven freezeout mechanisms \cite{Baker:2015qna,Choi:2018stw,Mohan:2019zrk,Belanger:2021smw,Belfatto:2021ats}. The interpretation of these searches exclude DM masses of about $0.1$--$1$ TeV and mediator masses of about $0.5$--$5$ TeV depending on the theoretical model.  \\

In the light of this current situation, an important question arises: what if DM only couples to the lepton sector? From the theoretical standpoint, there is  {\it a priori} no fundamental principle that can prevent DM from coupling to leptons only. This class of models has been proposed some time ago in ref. \cite{Liu:2013gba} and was widely studied in the literature \cite{Bai:2014osa,Chang:2014tea,Agrawal:2014ufa,Garny:2015wea,Jueid:2020yfj,Horigome:2021qof,Liu:2021mhn}. There are many interesting implications for these models. First, the scattering  of the DM off the nucleus is induced at the one-loop order and therefore these models can evade easily direct detection constraints even for model parameters of order $\mathcal{O}(1)$. Second, except for electron-philic scenarios, constraints from positron indirect detection searches are also not important since their annihilation is dominated by $p$-wave amplitudes which are suppressed by the square of the DM velocity. Finally, the DM can be produced at the LHC through the decay of charged scalars and therefore the corresponding bounds are not as strong as in the case of mono--X searches  especially in the case of $SU(2)_L$ gauge singlet mediators \cite{Jueid:2020yfj}.  Therefore an efficient probe of this category of models is through leptonic colliders such as the International Linear Collider (ILC), Chinese Electron Positron Collider (CEPC), and the future muon colliders. Recently, future muon colliders are attracting high interest due to their capability to probe new physics beyond the SM at very high scales \cite{Delahaye:2019omf,Long:2020wfp,AlAli:2021let} and therefore competing with the future circular colliders (FCC--hh). On the other hand, these machines can achieve very high energies thanks to the expected excellent cooling systems and the weaker synchrotron radiation. Finally at very high energies, muon colliders are necessarily vector-boson colliders where the dominant production channels are through vector-boson fusion (VBF) \cite{Costantini:2020stv,Ruiz:2021tdt}. Phenomenology of both the SM and beyond at muon colliders has been extensively studied in the literature (see {\it e.g.} \cite{Capdevilla:2020qel,Chiesa:2020awd,Han:2020uid,Han:2020uak,Yin:2020afe,Huang:2021nkl,Capdevilla:2021rwo,Capdevilla:2021fmj,Asadi:2021gah,Casarsa:2021rud,Liu:2021akf,Han:2021udl,Han:2021kes,Han:2021lnp,Lv:2022pts,Liu:2022byu,Azatov:2022itm,Yang:2022fhw,Bao:2022onq,Chen:2022msz} and references therein).  \\ 
  
In this work, we study the production of DM at muon colliders within the minimal lepton portal DM model in which we extends the SM with two $SU(2)_L$ singlets: a charged scalar that plays the role of the mediator and a neutral right-handed fermion (or, equivalently, Majorana particle) that plays the role of the DM candidate. We first comprehensively the impact of the different theoretical and experimental constraints on the model parameter space in the muon-philic scenario, {\it i.e.} the scenario where the DM couples predominantly to muons. We then select a few benchmark points that define phenomenologically viable scenarios that can be probed at high energy muon colliders. We study the production cross sections and the expected backgrounds for a set of production channels totaling 26 production channels for DM and 28 production channels for the charged singlet scalar. A particular feature of this model is that the DM is a Majorana fermion and therefore does not couple to gauge bosons directly and therefore the direct production of DM does not receive any contribution from VBF channels. We select a few production channels that can have high discovery potential and discuss the possible signatures and the associated backgrounds. This work is an introduction for future projects where a complete exploration of the model at muon colliders will be performed. \\

The remainder of this paper is orgnized  as follows. We discuss the model and its UV completion in section \ref{sec:model} along with the  constraints from LEP searches, $H_{\rm SM} \to \gamma\gamma$ and theoretical constraints. In section \ref{sec:LFV} we discuss the constraints from charged lepton flavour violation in $\ell_\alpha \to \ell_\beta \gamma$, $\ell_\alpha \to 3 \ell_\beta$ and $H_{\rm SM} \to \ell_\alpha \bar{\ell}_\beta$.  A detailed analysis of DM phenomenology in this  model is presented in section \ref{sec:DM} where we discuss the DM relic density, direct detection constraints and Higgs invisible decays. A study of DM production at muon colliders, the interesting signatures and the associated backgrounds is performed in section \ref{sec:prod:NR}. In section \ref{sec:prod:S} we study the production of charged scalars at muon colliders. We draw our conclusions in section \ref{sec:conclusions}. 

\section{Theoretical setup}
\label{sec:model}

\subsection{The model}

We consider a minimal extension of the SM by two gauge singlet fields: a charged scalar ($S$) and a right-handed fermion ($N_R$). 
We further assume that the two extra singlets are odd under $Z_2$ symmetry while all the SM particles are even; {\it i.e.,} $\{S, N_R\} \to \{-S, -N_R\}$ and $\{\ell, q, \nu, \Phi, V^\mu\} \to \{\ell, q, \nu, \Phi, V^\mu\}$. To ensure that the $N_R$ state is a suitable DM candidate within our model, we impose the condition $M_{N_R} < M_S$. Furthermore, the charged singlet is assumed to carry a lepton number and therefore couples only to charged leptons.\footnote{This charged singlet is also called a scalar lepton \cite{Baum:2020gjj} and the relevant interaction Lagrangian is similar to the case of interaction of supersymmetric slepton with a neutralino and a charged lepton. The difference here is that we assume a single charged scalar to couple to all the leptons instead of three scalars, usually denoted by $\tilde{e}_R, \tilde{\mu}_R,~{\rm and}~ \tilde{\tau}_R$, where each scalar couples to a specific lepton generation.} The full Lagrangian is given by
\begin{eqnarray}
\mathcal{L} = \mathcal{L}_{\rm SM} + \mathcal{L}_{S} - V(\Phi, S),
\end{eqnarray}
where $\Phi$ refers to the SM Higgs doublet, $\mathcal{L}_S$ is the interaction Lagrangian for the singlet scalar (including the kinetic term),  and $V(\Phi, S)$ is the scalar potential. 
The interaction Lagrangian for the $S$ field is given by
\begin{eqnarray}
\mathcal{L}_S = \sum_{\ell=e,\mu,\tau} Y_{\ell N} \overline{\ell}^c_R S N_R + (\mathcal{D}^\mu S)^\dagger(\mathcal{D}_\mu S) + {\rm h.c.},
\label{eq:lag:S}
\end{eqnarray}
with $\mathcal{D}_\mu S = (\partial_\mu - i g_2 Y_S B_\mu/2) S$ being the covariant derivative, $Y_S = 2$ is the hypercharge of the scalar singlet and $g_2$ is the $U(1)_Y$ gauge coupling. The kinetic term in equation \eqref{eq:lag:S} gives rise to interaction with $A_\mu$ and $Z_\mu$ which are given, after field rotations, by
\begin{eqnarray*}
\mathcal{L}_{S; \rm gauge} &=& -(e A^\mu - e \tan\theta_W Z^\mu) S^\dagger \bar{\partial}_\mu S + 
 e^2 A_\mu A^\mu S^\dagger S \nonumber \\ &+& e^2 \tan^2\theta_W Z_\mu Z^\mu S^\dagger S - 2 e^2 \tan\theta_W A_\mu Z^\mu S^\dagger S, 
\end{eqnarray*}
where $e = \sqrt{4 \pi \alpha_{\rm EM}}$ is the electric charge, $\theta_W$ is the Weinberg mixing angle, and $A\bar{\partial_\mu}B \equiv A(\partial_\mu B) - (\partial_\mu A)B$. 
The most general {\it CP}-conserving, renormalizable and gauge invariant scalar potential is given by 
\begin{widetext}
\begin{equation}
V(\Phi, S) = - M_{11}^2 |\Phi^\dagger \Phi| + M_{22}^2 |S^\dagger S| + \lambda_1 |\Phi^\dagger \Phi|^2 + \lambda_2 |S^\dagger S|^2 + \lambda_3 |\Phi^\dagger \Phi| |S^\dagger S|.
\end{equation}
\end{widetext}
All the parameters of the scalar potential are assumed to be real valued as a consequence of {\it CP} conservation. The process of electroweak symmetry breaking leads to three physical scalars: $H_{\rm SM}$ identified with the recently discovered $125~{\rm GeV}$ SM Higgs boson and a pair of charged scalars denoted by $H^\pm$. Their masses are given at the lowest order in perturbation theory by
\begin{eqnarray}
M_{H_{\rm SM}}^2 = \lambda_1 \upsilon^2 = - 2 M_{11}^2, \quad M_{H^\pm}^2 = M_{22}^2 + \frac{1}{2} \lambda_3 \upsilon^2,
\end{eqnarray}
with $\upsilon$ being the vacuum expectation value (VEV) of the SM Higgs doublet. This model involves seven additional free parameters which we parametrise as follows
\begin{eqnarray}
\{M_{H^\pm}, M_{N_R}, \lambda_2, \lambda_3, Y_{e N}, Y_{\mu N}, Y_{\tau N} \}.
\label{eq:params}
\end{eqnarray}
For convenience we define the combination of the couplings $Y_{\ell N}$ by\footnote{This is equivalent to a definition of a system of spherical coordinates wherein the new parameters are $Y_{\ell N}, \theta$ and $\varphi$ such that $\theta \in [0, \pi]$ and $\varphi \in [0, 2\pi]$. The couplings in equation \eqref{eq:lag:S} are defined here as $Y_{e N} = Y_{\ell N} \cos\varphi\sin\theta$, $Y_{\mu N} = Y_{\ell N} \sin\varphi\sin\theta$  and $Y_{\tau N} = Y_{\ell N} \cos\theta$.}
$$
Y_{\ell N} = \sqrt{Y_{e N}^2 + Y_{\mu N}^2 + Y_{\tau N}^2},
$$
which is a very good parametrisation in case the charged leptons are assumed to be massless.

\subsection{Theoretical and experimental constraints}
The parameters of the model in equation \eqref{eq:params} are subject to various theoretical and experimental constraints. We start with  a brief discussion of the constraints influencing the scalar potential parameters and $M_{H^\pm}$ where more details can be found in \cite{Jueid:2020yfj}. The width of the SM Higgs boson is only affected by the rate of its decay to $\gamma\gamma$. In this model, this process receives new contributions from charged singlet scalar which give rise to destructive or constructive contributions depending on the sign of $\lambda_3$ \cite{Swiezewska:2012eh, Arhrib:2012ia, Jueid:2020rek}\footnote{We have found a typo in the analytical expression in ref.  \cite{Arhrib:2012ia} which may influences their numerical results.}. In the present work, we have used the most recent ATLAS-CMS combined measurement of $|\kappa_\gamma|$ \cite{Khachatryan:2016vau}
$$ 
|\kappa_\gamma| \equiv \sqrt{\Gamma(H\to\gamma\gamma)/\Gamma(H\to\gamma\gamma)_\mathrm{SM}} = 0.87^{+0.14}_{-0.09}.
$$
We assume the theoretical prediction to be in agreement with the experimental measurement at the $2\sigma$--level.  We found that the enhancement of $|\kappa_\gamma|$ always occur for $\lambda_3 < 0$ which excludes charged scalars with masses up to $\sim 380~{\rm GeV}$ \cite{Jueid:2020rek}. For $\lambda_3 > 0$, we get three possible regimes: {\it (i)} large and negative contribution that implies an enhancement of $\kappa_\gamma$, {\it (ii)} positive but small contribution which makes $\kappa_\gamma$ consistent with the experimental measurement and {\it (iii)} exact or almost exact cancellation between the $H^\pm$ and the $W$-boson contributions which make $\kappa_\gamma$ very small. Therefore, for $\lambda_3 > 0$, charged singlet masses up to $380~{\rm GeV}$ are excluded but with small region where the constraints completely vanish. \\

In addition to constraints from Higgs decays, the parameters of the scalar potential are subject to a number of theoretical constraints. We note that the bounds on the scalar potential of this model can be obtained from those in {\it e.g.} the inert doublet model by setting $\lambda_4 = \lambda_5 = 0$. In this study, we impose constraints from vacuum stability conditions (or boundness-from-below) \cite{Branco:2011iw}, perturbativity, perturbative unitarity \cite{Kanemura:1993hm, Akeroyd:2000wc} and false vacuum \cite{Ginzburg:2010wa}. The false vacuum condition plays a very important role in constraining the parameters $\lambda_2, \lambda_3$ and $M_{H^\pm}$. We get 
\begin{eqnarray}
M_{H^\pm}^2 \geqslant \frac{1}{2} \bigg(\lambda_3 \upsilon^2 - M_{H_{\rm SM}}^2 \sqrt{\frac{\lambda_2}{\lambda_1}} \bigg).
\end{eqnarray}
We found that: {\it (i)} $\lambda_3$ cannot be larger $5$ for all charged scalar masses and {\it (ii)} there is a parabola in the plane defined by $\lambda_3$ and $M_{H^\pm}$ which simply tells us that the smaller is the minimum allowed value of $M_{H^\pm}$ the smaller is the maximum allowed value of $\lambda_3$. These conclusions are mildly dependent on the choice of $\lambda_2$ and, therefore, we choose $\lambda_2 = 2$ in the remainder of this manuscript without loss of generality. \\

\begin{figure}[!t]
    \centering
    \includegraphics[width=0.9\linewidth]{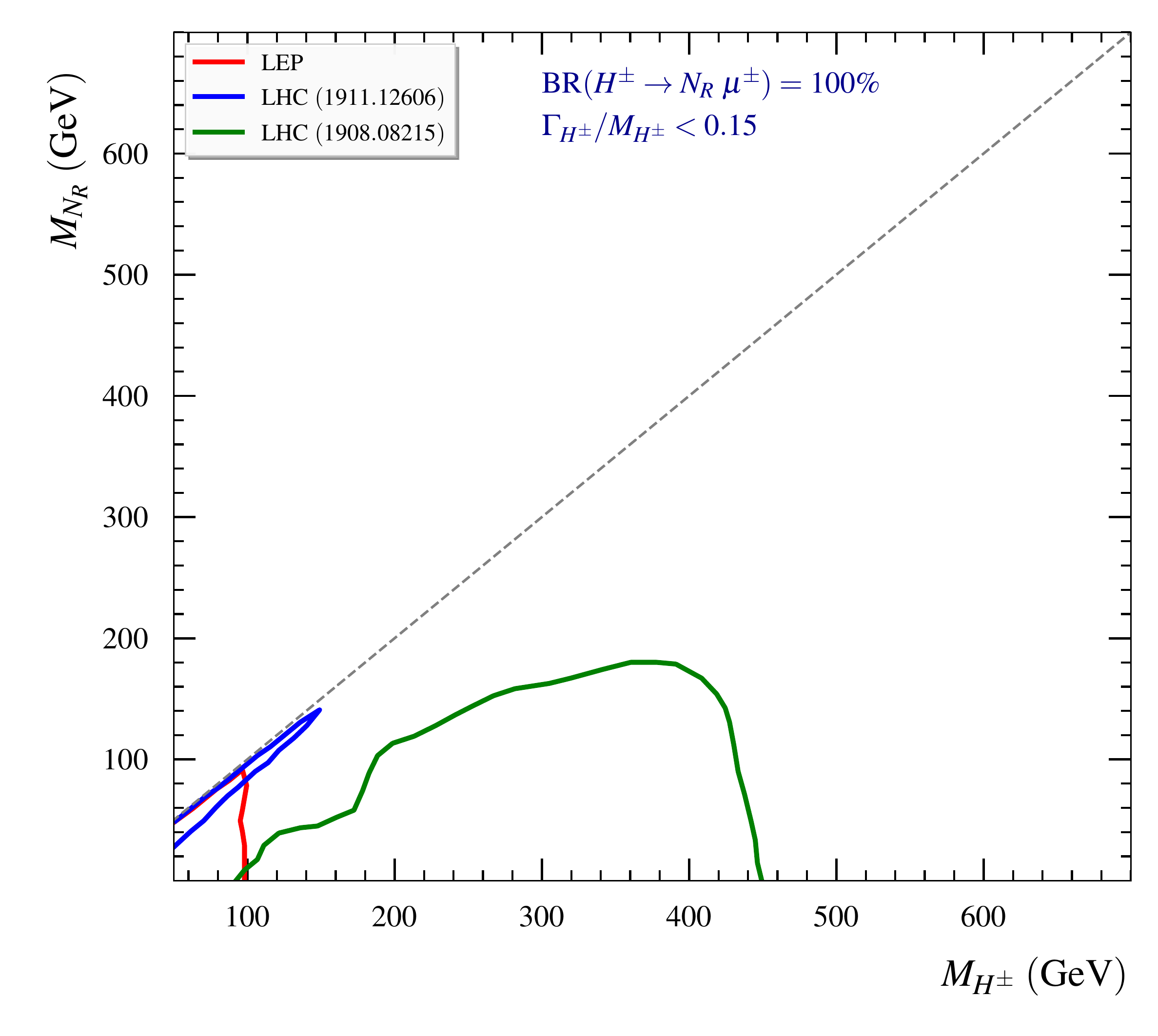}
    \vspace{-0.2cm}
    \caption{Summary of the collider constraints on the parameter space of the model displayed on the plane of $(M_{H^\pm}, M_{N_R})$. We show the constraints from LEP searches of sleptons and charginos (red), LHC searches for sleptons in the compressed regime (blue) and constraints from LHC searches of sleptons and charginos for large mass splittings (green). Here, we assume that the charged singlet scalar decays to $\mu^\pm N_R$ with a branching fraction of $100\%$ and assume the Narrow Width Approximation (NWA) by selecting parameters for which we have $\Gamma_{H^\pm}/M_{H^\pm} < 0.15$. The gray dashed line corresponds to the kinematical boundary above which the $N_R$ particle is not a suitable dark matter candidate.}
    \label{fig:collider:bounds}
\end{figure}

The model can be constrained by using the null results of LEP and LHC searches for supersymmetric particles \cite{Abbiendi:2003ji,ATLAS:2019lff, ATLAS:2019lng}. The OPAL collaboration of the LEP experiment has searched for charginos decaying into a charged lepton and the lightest supersymmetric neutralino using $680~{\rm pb}^{-1}$ of integrated luminosity \cite{Abbiendi:2003ji}. Assuming that the branching ratio of $H^\pm \to \mu^\pm N_R$ is $100\%$, the production of charged singlet pairs occurs through gauge interactions ($s$--channel diagrams with the exchange of $\gamma^*/Z^0$). This search constrain the mass of the charged singlet to be not heavier than $100~{\rm GeV}$ for any value of $Y_{\mu N}$. This can be seen clearly in the red contour of figure \ref{fig:collider:bounds}. The ATLAS collaboration at the LHC has also searched for sleptons and charginos assuming $100\%$ branching fraction to a charged lepton and neutralino. These searches targeted large mass splitting $\Delta = m_{\tilde{\ell}} - m_{\chi^0} \geq 80~{\rm GeV}$ \cite{ATLAS:2019lff} and compressed spectra for a mass splitting as low as $0.55~{\rm GeV}$ \cite{ATLAS:2019lng}. The two searches utilized a total luminosity of $139~{\rm fb}^{-1}$. The first search constrained scalar singlet masses to be lighter than about $440$ GeV while the search for compressed spectra constrain the whole compressed region for $M_{H^\pm}$ up to $150$ GeV (see the blue and green contours of figure \ref{fig:collider:bounds}).   

\subsection{Examples of UV completions}
In this section, we discuss the UV completions of this minimal framework. In general, there are two ways to UV complete the first term in $\mathcal{L}_S$: {\it (i)} assume it to be a part of a radiative neutrino mass model or {\it (ii)} embed it in a grand-unified theory -- $SU(5)$ for example --. We start by the radiative neutrino mass models. The most economical way to extend this model is through the so-called Krauss-Nasri-Trodden (KNT) three-loop radiative neutrino mass model \cite{Krauss:2002px}. In addition to $S$ and $N_R$, the KNT model extends the SM with an additional scalar singlet that is even under $Z_2$. Another possibility is through the so-called the scotogenic model which extends the SM with one inert doublet and three right-handed fermions \cite{Ma:2006km}. The phenomenology of the scotogenic model has been widely studied in the literature \cite{Ahriche:2017iar, Baumholzer:2018sfb, Borah:2018rca, Kitabayashi:2018bye, Ahriche:2018ger, Baumholzer:2019twf, Borah:2020wut,Sarazin:2021nwo}. The relevant interaction becomes 
\begin{eqnarray}
\mathcal{L} \supset h_{\alpha \beta} \bar{L}_{L \alpha} (i \sigma_2) \Phi_{\rm IDM} N_\beta \supset h_{\alpha \beta} \bar{\ell}_{L \alpha} S N_{\beta},
\label{eq:lag:IDM}
\end{eqnarray}
where $\Phi_{\rm IDM} = (S, (h_2+ia_2)/\sqrt{2})^T$, and $\alpha, \beta$ are generation indices. Identifying \eqref{eq:lag:IDM} with the first term in $\mathcal{L}_S$ we have $Y_{e N} = h_{11}, Y_{\mu N} = h_{21}, Y_{\tau N} = h_{31}$. We must stress out that the gauge interactions of the singlet scalar in this model are different from the scotogenic model due to the fact that $S$ is a member of $SU(2)_L$ doublet while it is a singlet in the present model. 

The first term in equation \eqref{eq:lag:S} can be obtained from a grand-unified theory; For example, by embeding the SM into a $SU(5)$ gauge group with the matter fields belonging to the ${\bf{10}}_{F}$ and $\bar{{\bf{5}}}_F$ representations,  the charged singlet belongs to the ${\bf{10}}_{H}$ representation, and the right handed neutrino belongs to the singlet representation ${\bf{1}}_{\alpha}$, which in this case we can write
\begin{eqnarray}
\mathcal{L}_{\text{int}}= g_{\alpha \beta} \overline{{\bf{10}}}_\alpha \otimes {\bf{10}}_{H} \otimes {\bf{1}}_{N_\beta}\supset g_{\alpha \beta} \ell_{R\alpha}^T C N_\beta S^+.
\end{eqnarray}

In addition to the minimal $SU(5)$, we can obtain the first term of equation \eqref{eq:lag:S} from a flipped-$SU(5) \otimes U(1)_X$ grand-unified theory. Here, the right-handed charged lepton field is a singlet under $SU(5)$ while the right-handed neutral fermion ($N_R$) is a member of the ${\bf{10}}_\alpha$ representation. In this case, we have
\begin{eqnarray}
\mathcal{L}_{\text{int}} &=& \frac{h_{\alpha \beta}}{\Lambda} \overline{{\bf{10}}}_\alpha \otimes \bar{{\bf{1}}}_\beta \otimes {\bf{10}}_{H} \otimes {\bf{1}}_{S} + h.c. \nonumber \\ 
&\supset& \frac{h_{\alpha\beta} \langle {\bf{10}}_H \rangle}{\Lambda} N^T C \ell_R S^-,
\end{eqnarray}
where we integrated out a heavy intermediate state with a scale $\Lambda \gg \Lambda_{\rm GUT}$. \\

\section{Charged Lepton Flavour violation}
\label{sec:LFV}

The interaction Lagrangian in equation \eqref{eq:lag:S} conserves total lepton number to all orders in perturbation theory since the charged singlet possesses a lepton number\footnote{To generate a Majorana neutrino mass one has extend the Lagrangian \eqref{eq:lag:S} so that the total lepton number is violated. The minimal realization of such breaking can be achieved by having, in addition to $S$, a second  $SU(2)$ singlet charged scalar with lepton number equals to two units; which is the KNT model.}. However, the charged singlet scalar can give rise to processes violating  flavor lepton numbers $L_\alpha; \alpha=e, \mu, \tau$ at the one-loop order. These processes called charged lepton flavor violating (CLFV) processes are categorised into three categories: ({\it i}) $\ell_\alpha = \ell_\beta \gamma$, ({\it ii}) $\ell_\alpha \to \ell_\beta \ell_\beta \bar{\ell}_\beta$ and ({\it iii}) $e$--$\mu$ conversion in nuclei. In this section we discuss the impact of the CLFV constraints on the model parameter space. The most stringent bounds on the couplings $Y_{\ell_\alpha N}$ come from the branching ratio of $\mu \to e \gamma$ decay. The analysis of the CLFV decays in this work are heavily based on the results of refs. \cite{Hisano:1995cp, Arganda:2005ji, Ilakovac:2012sh, Toma:2013zsa}. A summary of the current and future bounds on the CLFV decays is shown in Table \ref{tab:LFV:bounds}. 

\begin{table}[tb!]
\centering
\begin{tabular}{l  c  c}
\toprule
CLFV decay & \hspace{0.5cm} Present limit \hspace{0.5cm} & Future sensitivity  \\
\toprule
$\mu \to e \gamma$ & $5.7 \times 10^{-13}$ \cite{Adam:2013mnn} & $6 \times 10^{-14}$ \cite{Baldini:2013ke}  \\
$\tau \to e \gamma$ & $3.3 \times 10^{-8}$ \cite{Aubert:2009ag}& $\sim 10^{-8}-10^{-9}$ \cite{Hayasaka:2013dsa}\\
$\tau \to \mu \gamma$ & $4.4 \times 10^{-8}$ \cite{Aubert:2009ag}& $\sim 10^{-8}-10^{-9}$ \cite{Hayasaka:2013dsa} \\
$\mu \to eee$ & $1.0 \times 10^{-12}$~\cite{SINDRUM:1987nra} & $\sim 10^{-16}$ \cite{Blondel:2013ia}\\
$\tau \to eee$ & $2.7\times10^{-8}$~\cite{Hayasaka:2010np} & $\sim 10^{-9}-10^{-10}$ \cite{Hayasaka:2013dsa}  \\
$\tau \to \mu\mu\mu$ & $2.1\times10^{-8}$~\cite{Hayasaka:2010np} & $\sim 10^{-9}-10^{-10}$ \cite{Hayasaka:2013dsa}  \\
$H_{\rm SM} \to \mu \tau$ & $1.5 \times 10^{-3}$ \cite{CMS:2021rsq} & $-$ \\
$H_{\rm SM} \to e\tau$    & $2.2 \times 10^{-3}$ \cite{CMS:2021rsq} & $-$ \\
$H_{\rm SM} \to e\mu$     & $3.5 \times 10^{-4}$ \cite{CMS:2016cvq} & $-$ \\
\bottomrule
\end{tabular}
\caption{Current experimental bounds and future sensitivities for low-energy CLFV decays and high-energy Higgs boson LFV decays.}
\label{tab:LFV:bounds}
\end{table}

\subsection{$\ell_\alpha \to \ell_\beta \gamma$}

The radiative decays of charged leptons ($\ell_\alpha \to \ell_\beta \gamma$) receive contributions from the exchange of the charged singlet scalar and Majorana DM. After computing the one-loop integrals we get the effective magnetic dipole operator $\mu^M_{\beta\alpha} \overline{\ell}_\beta\sigma^{\mu\nu} \ell_\alpha F_{\mu\nu}/2$ with $\mu_{\beta\alpha} = e m_\alpha A_{\rm M}/2$ and $A_{\rm M}$ is given by
$$
A_{\rm M} = \frac{Y_{\ell_\beta N}Y_{\ell_\alpha N}}{2(4\pi)^2} \frac{1}{M_{H^\pm}^2} \mathcal{F}(\xi), 
$$
where $\xi = M_{N_R}^2/M_{H^\pm}^2$ and $\mathcal{F}(x) = (1 - 6 x + 3 x^2 + 2 x^3 - 6 x^2 \log x)/(6(1-x)^4$ is the one-loop function which have the following limits $\mathcal{F}(x) \to 1/6~(1/12)$ for $x\to 0~(1)$.  The resulting decay branching ratio can be computed easily to give
\begin{widetext}
\begin{equation}
{\rm BR}\left(\ell_{\alpha}\to\ell_{\beta}\gamma\right)=
\frac{3(4\pi)^3 \alpha_{\rm EM}}{4G_F^2} 
|A_{\rm M}|^2 \times 
{\rm BR}\left(\ell_{\alpha}\to\ell_{\beta}\nu_{\alpha}.
\overline{\nu}_{\beta}\right), \hspace{0.5cm}
\label{eq:LFV:ltola}
\end{equation}
\end{widetext}
Here, $G_F = 1.166 \times 10^{-5}~{\rm GeV}^{-2}$, $\alpha_{\rm EM} = 1/137$, and ${\rm BR}\left(\ell_{\alpha}\to\ell_{\beta}\nu_{\alpha}
\overline{\nu_{\beta}}\right)$ is the SM decay branching ratios. We choose ${\rm BR}(\mu\to e\nu\bar{\nu}), {\rm BR}(\tau \to e\nu\bar{\nu}), {\rm BR}(\tau\to \mu\nu\bar{\nu}) \approx 1, 0.1783, 0.1741$ \cite{ParticleDataGroup:2018ovx}.

\begin{figure}[!t]
    \centering
    \includegraphics[width=0.93\linewidth]{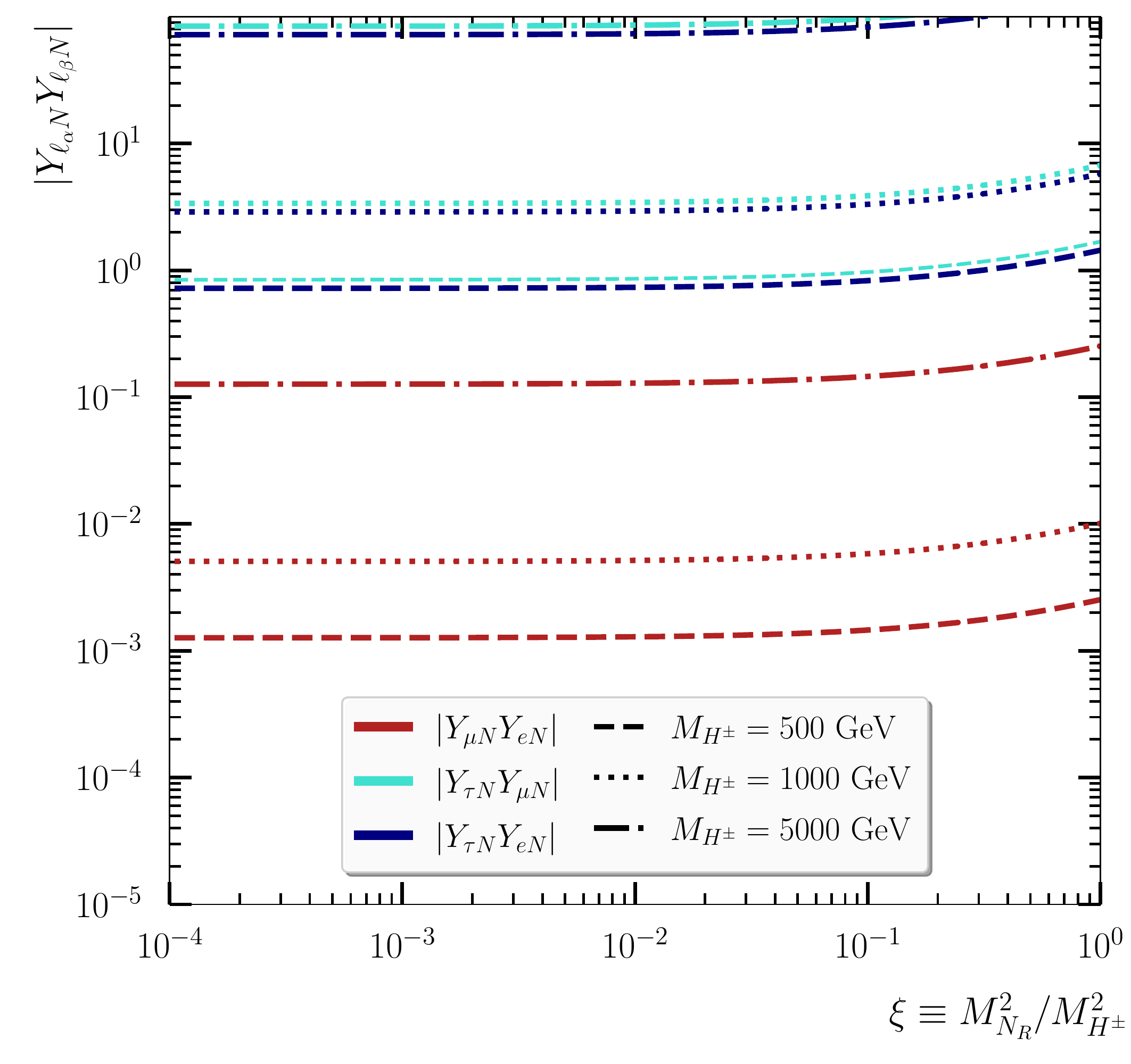}
    \caption{The maximum value of the products of the $Y_{\ell_\alpha N} Y_{\ell_\beta N}$ as a function of $\xi$ for different values of the charged singlet mass $M_{H^\pm}$. The results are shown for $M_{H^\pm} = 500~{\rm GeV}$ (dashed), $M_{H^\pm} = 1000~{\rm GeV}$ (dotted) and $M_{H^\pm} = 5000~{\rm GeV}$ (dash-dotted).}
    \label{fig:LFV:bounds}
\end{figure}

Using the most recent experimental bounds on ${\rm BR}(\ell_\alpha \to \ell_\beta \gamma)$ from the \textsc{Meg} \cite{Adam:2013mnn} and \textsc{BaBar} \cite{Aubert:2009ag} experiments, we can use equation \eqref{eq:LFV:ltola} to derive the following bounds on the products of the couplings: 
\begin{eqnarray}
|Y_{e N} Y_{\mu N}| < \bigg(\frac{2.855 \times10^{-5}}{\mathrm{GeV}} \bigg)^2 \frac{M_{H^\pm}^2}{|\mathcal{F}(\xi)|}, \nonumber \\
|Y_{e N} Y_{\tau N}| < \bigg(\frac{4.428 \times10^{-4}}{\mathrm{GeV}} \bigg)^2 \frac{M_{H^\pm}^2}{|\mathcal{F}(\xi)|}, \\
|Y_{\tau N} Y_{\mu N}| < \bigg(\frac{4.759 \times10^{-4}}{\mathrm{GeV}} \bigg)^2 \frac{M_{H^\pm}^2}{|\mathcal{F}(\xi)|}. \nonumber
\end{eqnarray}
Since the one-loop function varies roughly between $1/12$ and $1/6$, the upper bound on the coupling $Y_{\ell_\alpha N} Y_{\ell_\beta N}$ is proportional to the square of the charged singlet mass with almost no dependence on $M_{N_R}$. Therefore, limits are expected to be strong for light $H^\pm$ and become very weak for heavy $H^\pm$. This can be clearly seen in figure \ref{fig:LFV:bounds} where the maximum allowed values of $|Y_{\ell_\alpha N} Y_{\ell_\beta} N|$ by the CLFV decays ${\rm BR}(\ell_\alpha \to \ell_\beta \gamma)$ are shown as a function of $\xi = M_{N_R}^2/M_{H^\pm}^2$ for $M_{H^\pm} = 500, 1000,~{\rm and}~5000~{\rm GeV}$. As expected, the bounds on $|Y_{eN} Y_{\mu N}|$ are the strongest ones while the bounds on $|Y_{\tau N} Y_{\mu N}|$ and $|Y_{e N} Y_{\tau N}$ are similar. We must stress out that the CLFV decays do not constraints the $Y_{\ell_\alpha N}$ couplings  {\it per se} but only their products. Following this finding, there is some freedom regarding the choice of the couplings which we call here benchmark scenarios (see next sections). Given that this study is mainly concerned about the phenomenology of the leptophilic DM models at muon colliders, we choose a scenario where the coupling of dark matter to the muon is quite large while the other couplings are chosen such that they fulfill the experimental bounds on CLFV decays: $Y_{\mu N} \simeq \mathcal{O}(1) \gtrsim Y_{\tau N} \gg Y_{e N}$. \\

\subsection{$\ell_\alpha \to \ell_\beta \ell_\beta \bar{\ell}_\beta$}

It is noteworthy to discuss the constraints from the CLFV decays $\ell_\alpha \to \ell_\beta \ell_\beta \bar{\ell}_\beta$. These processes receive four contributions at the one-loop order: penguin diagrams with the exchange of $\gamma$, $Z$ and $H_{\rm SM}$ and box diagrams. The contribution of the SM Higgs boson is suppressed due to the smallness of the Higgs-lepton Yukawa coupling. The corresponding branching ratio is given by \cite{Toma:2013zsa}
\begin{widetext}
\begin{eqnarray}
{\rm BR}(\ell_\alpha \to \ell_\beta \ell_\beta \bar{\ell_\beta}) &=& \frac{3(4\pi)^2\alpha_{\rm EM}}{8 G_F^2} \bigg[\overbrace{|A_{\rm ND}|^2 + |A_{\rm M}|^2 \left(\frac{16}{3} \log\left(\frac{m_\alpha}{m_\beta}\right) - \frac{22}{3}\right)}^{\gamma~{\rm penguin}} +  \overbrace{\frac{1}{3} (2 |Z_{\rm RR}|^2 + |Z_{\rm RL}|^2)}^{Z~{\rm penguin}} \nonumber \\
&+& \frac{1}{6} |B_{\rm box}|^2 +  \underbrace{2~{\rm Re}\left(-2 A_{\rm ND} A_{\rm M}^* + \frac{1}{3} A_{\rm ND} B_{\rm box}^* - \frac{2}{3} A_{\rm M} B_{\rm box}^*\right)}_{{\rm Interference}} \bigg] \times \mathcal{B}, 
\label{eq:LFV:lto3l}
\end{eqnarray}
\end{widetext}
where $\mathcal{B} \equiv {\rm BR}(\ell_\alpha \to \ell_\beta \nu_\alpha \bar{\nu}_\beta)$. The contribution of the $\gamma$-penguins consist of the magnetic or dipole ($A_{\rm M}$) and the non-dipole ($A_{\rm ND}$) contributions. The dipole contribution is the same as of ${\rm BR}(\ell_\beta \to \ell_\beta\gamma)$ but enhanced by a factor of $16\times( \log(m_\alpha/m_\beta) - 22)/3$ which varies between $7$ and $36$ for $\tau\to 3\mu$ and $\tau \to 3 e$ respectively. The non-dipole contribution is given by
$$
A_{\rm ND} = \frac{Y_{\ell_\alpha N} Y_{\ell_\beta N}}{6 (4\pi)^2} \frac{1}{M_{H^\pm}^2} \mathcal{G}(\xi), 
$$
with $\mathcal{G}(x) = (2 - 9x + 18 x^2 - 11 x^3 + 6 x^3 \log x)/(6(1-x)^4)$ being the one-loop function for the non-dipole $\gamma$--penguin. This function has the following limits: $\lim_{x\to 0}\mathcal{G}(x) = 1/3$ and $\lim_{x\to 1}\mathcal{G}(x) = 1/4$. Therefore, the dipole $\gamma$-penguin contribution is large as compared to the non-dipole contributions; $\lim_{x\to 0}~(\lim_{x\to 1}) A_{\rm M}/A_{\rm ND}\times (16/3 \log(m_\alpha/m_\beta) - 22/3)  \approx \{3.5, 11, 18\}~(\{2, 7, 12\})$ for $\tau\to 3\mu, \mu \to 3 e,~{\rm and}~\tau \to 3 e$ respectively. The $Z$-penguin contribution is given by 
\begin{eqnarray}
Z_{\rm RR} = \frac{g_R^\ell Z_{\rm ND}}{g_1^2 \sin^2\theta_W M_Z^2},~Z_{\rm RL} = \frac{g_L^\ell Z_{\rm ND}}{g_1^2 \sin^2\theta_W M_Z^2},
\end{eqnarray}
where $g_R^\ell, g_L^\ell$ are the right and left-handed components of the $Z$-boson couplings to charged leptons, $g_1$ is the $SU(2)_L$ gauge coupling, $\sin\theta_W$ is the sine of the Weinberg mixing angle, and $Z_{\rm ND}$ is the momentum-independent $Z$-boson form factor which is given by
$$
Z_{\rm ND} = \frac{Y_{\ell_\alpha N} Y_{\ell_\beta N}}{2(4\pi)^2} \frac{m_\alpha m_\beta}{M_{H^\pm}^2} \frac{g_1}{\cos\theta_W} \mathcal{F}(\xi).
$$
We can see that the $Z$-penguin contribution involves an extra suppression by a factor of $m_\alpha m_\beta$ as compared to the dipole $\gamma$-contribution. Finally, the box contribution is given by
\begin{eqnarray}
B_{\rm box} = \frac{Y_{\ell_\alpha N} Y_{\ell_\beta N}^3}{2^7 \pi^3 \alpha_{\rm EM} M_{H^\pm}^2} \bigg[\mathcal{D}_1(\xi) + 2 \xi \mathcal{D}_2(\xi)\bigg],
\end{eqnarray}
where $\mathcal{D}_{1,2}(x)$ are the one-loop box functions given by $\mathcal{D}_1(x) = (-1 + x^2 - 2 x\log x)/(1-x)^3$ and $\mathcal{D}_2(x) = (-2 + 2 x - (1+x) \log x)/(1-x)^3$. The contribution of the box diagrams, contrarily to penguins, has an extra factor of $Y_{\ell_\beta N}^2$. Therefore, it may dominate for large couplings of the daughter lepton to DM. In this work, we check that the benchmark scenarios satisfy the bounds from the $\ell_\alpha \to 3\ell_\beta$ decays (see Table \ref{tab:BSs}). \\

\subsection{$H_{\rm SM} \to \ell_\alpha \bar{\ell}_\beta$}

We close this section by a brief discussion of the CLFV decays of the SM Higgs boson. These decays have been searched for by the ATLAS and the CMS collaborations with the most strongest bounds are reported on by CMS collaboration \cite{CMS:2021rsq, CMS:2016cvq}. In this model, the CLFV decays of the SM Higgs boson are degenerate to the radiative CLFV decays of the charged leptons. The constraints from CLFV of charged leptons imply that the CLFV decays of the SM Higgs boson are extremely suppressed and may even be beyond the future reach of the LHC and future colliders. The SM Higgs boson decay into $\ell_\alpha \ell_\beta$ is given by \cite{Herrero-Garcia:2016uab} 
\begin{widetext}
\begin{eqnarray}
{\rm BR}(H_{\rm SM} \to \ell_\alpha \ell_\beta) \simeq 1.2 \times 10^3 \times |y_{\ell_\alpha} Y_{\ell_\alpha N} Y_{\ell_\beta N}|^2 \bigg(\frac{\lambda_3}{4\pi}\bigg)^2 \bigg(\frac{\upsilon}{M_{H^\pm}}\bigg)^4,
\end{eqnarray}
\end{widetext}
with $y_{\ell_\alpha} = m_{\ell_\alpha}/(\sqrt{2} \upsilon)$ is the Higgs-lepton Yukawa coupling of the heavier lepton (chosen here to be $\ell_\alpha$). In this formula, the contribution of the lighter lepton is neglected. We expect the bounds from $H_{\rm SM} \to \ell_\alpha \bar{\ell}_\beta$ searches to be very weak. This can be clearly seen in from table \ref{tab:BSs} for the benchmark points we have used in this study.

\section{Dark Matter}
\label{sec:DM}
In this section, we discuss the DM phenomenology within this model. We start with the  calculation of  the relic density of the$N_R$ particles in section \ref{sec:relic} and then move to a detailed analysis of the spin-independent DM-nucleus scattering cross section in section \ref{sec:direct}. Next, we derive the  constraint on the couplings $Y_{lN}$ by analysing  the   the  Higgs invisible decays and conclude by a selection of the benchmark points that are compatible with all the theoretical and the experimental constraints in section \ref{BPs}. 

\subsection{Relic density}
\label{sec:relic}
\begin{figure*}[!tbp]
\centering
\includegraphics[width=0.62\linewidth]{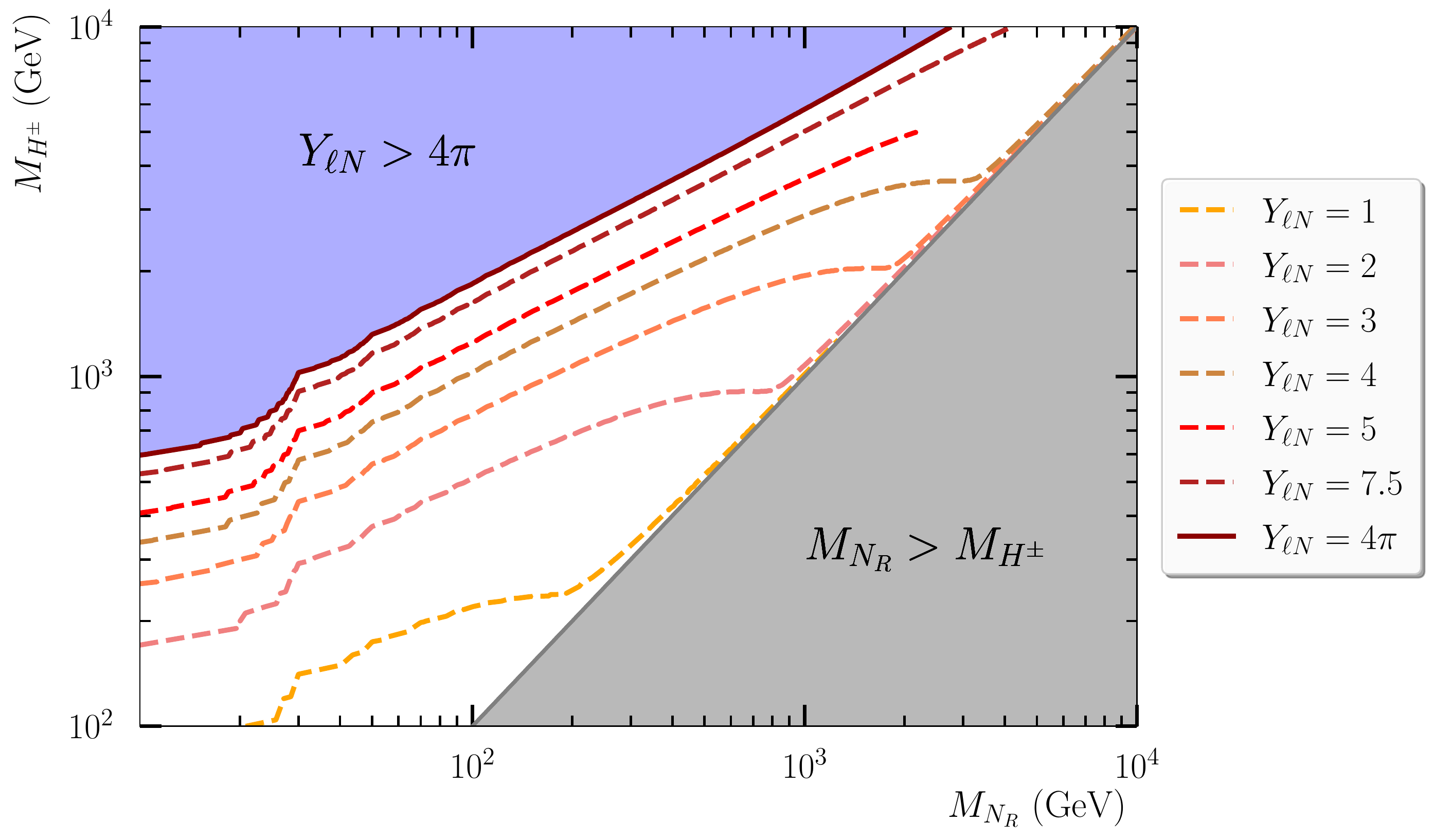}
\caption{Values of the coupling $Y_{\ell N}$ consistent with the measurement of the relic density by the \textsc{Planck} collaboration projected on the mass of the dark matter and the mass of the charged singlet scalar. The isolines corresponding to $\Omega h^2 \approx 0.12$ are shown for $Y_{\ell N} = 1, 2, 3, 4, 5, 7.5$ and $4\pi$. The blue shaded area corresponds to the region where the perturbativity is broken while the shaded gray region correspond to the kinematically forbidden region $M_{N_R} > M_{H^\pm}$ in which $N_R$ is not stable and therefore not a suitable dark matter candidate.}
\label{fig:relicdensity}
\end{figure*}

The relic density of the $N_R$ particles receives contributions from both the annihilation and the co-annihilation. The co-annihilation becomes active
when the mass splitting $\Delta \equiv M_{H^\pm} - M_{N_R} < 0.1 \times M_{N_R}$ while the annihilation contributes for the whole parameter space. 
For the annihilation, there are two major contributions: {\it (i)} $N_R N_R \to \ell_\alpha^+ \ell_\beta^-$ from the exchange of the charged
scalar singlet in $t$- and $u$-channels, and {\it (ii)} $N_R N_R \to \sum_{X \in {\rm SM}} X \overline{X}$ 
which arises from the exchange of the SM Higgs boson via $s$-channel diagrams. Note that $s$--channel contributions to the relic density are negligible 
in our model if one demands perturbativity of the couplings. The reason is that the leading order contribution to the $s$-channel annihilation amplitudes 
arises at the one-loop order. To obtain the relic density of the $N_R$ particles, one must solve the Boltzmann equations given by \cite{Scherrer:1985zt, Griest:1990kh, Jungman:1995df} 
\begin{eqnarray}
\frac{{\rm d}n_{N_R}}{{\rm d}t} + 3 H n_{N_R} = - 2 \langle \sigma_{N_R} v \rangle \bigg[(n_{N_R})^2 - (n_{N_R}^{\rm eq})^2\bigg],
\label{eq:boltzmann:ann}
\end{eqnarray}
with $H=\dot{a}/a$, $n_{N_R}$ is the number density of the $N_R$ particle and $n_{N_R}^{\rm eq} \approx g_{N_R} \left(\frac{M_{N_R} T}{2\pi}\right)^{3/2} e^{-M_{N_R}/T}$ is its number density at the thermal equilibrium. Note that in the absence of interactions that change the 
number density of $N_R$, the right handed side of equation \eqref{eq:boltzmann:ann} would be equal to zero and $n_{N_R} \propto a^{-3}$. This equation can be solved to 
give approximately 
\begin{eqnarray}
 \Omega_{\rm DM} h^2 \simeq \frac{3 \times 10^{-27}~{\rm cm}^3 {\rm s}^{-1}}{\langle \sigma(x_f) v \rangle}, 
\end{eqnarray}
where $\langle \sigma(x_f) v \rangle$ is the thermally-averaged annihilation cross section for the $N_R$ particle
\begin{widetext}
\begin{eqnarray}
\langle \sigma(x_f) v \rangle = \frac{1}{8 M_{N_R}^4 T_f K_2^2(M_{N_R}/T_f)} \sum_{\alpha, \beta} \int_{4 M_{N_R}^2}^{\infty} {\rm d}\hat{s} \sqrt{\hat{s} - 4 M_{N_R}^2} K_1(\sqrt{\hat{s}}/T_f)\sigma_{N_R N_R \to \ell_\alpha \ell_\beta}(\hat{s}),
\end{eqnarray}
\end{widetext}
where $K_1(x)$ and $K_2(x)$ are the modified Bessel functions of the second kind and $\sigma_{N_R N_R \to \ell_\alpha \ell_\beta}(\hat{s})$ is the annihilation cross section into charged lepton which is given by
\begin{widetext}
\begin{equation}
\sigma_{N_R N_R \to \ell_\alpha \ell_\beta}(\hat{s}) = \frac{1}{2^3\pi} \frac{|Y_{\ell_\alpha N} Y_{\ell_\beta N}|^2}{\hat{s}~\hat{\kappa}_1} \bigg[(m_{\ell_\alpha}^2 + m_{\ell_\beta}^2) (\hat{s} - 2 M_{N_R}^2) + \frac{1}{6} \frac{\hat{\kappa}_2}{\hat{\kappa}_1} \hat{s}(\hat{s} - 4 M_{N_R}^2)\bigg],
\end{equation}
\end{widetext}
where $\hat{\kappa}_i \equiv \hat{\kappa}_i(M_{H^\pm}^2, M_{N_R}^2, \hat{s})$,  $\hat{\kappa}_1(x,y,z) = (2 x + 2 y - z)^2$ and $\hat{\kappa}_2(x,y,z) = (4x - 4 y + z)^2 - 2 z^2$. To simplify the discussion about the relic density, we consider the annihilation cross section in the limit $\hat{s} \to 4 M_{N_R}^2$
\begin{eqnarray*}
\sigma_{N_R N_R \to \ell_\alpha \ell_\beta} \approx \frac{|Y_{\ell_\alpha N} Y_{\ell_\beta N}|^2}{2^6 \pi M_{H^\pm}^4} (m_{\ell_\alpha}^2 + m_{\ell_\beta}^2)  \bigg(1 + \frac{M_{N_R}^2}{M_{H^\pm}^2}\bigg)^{-2}.
\end{eqnarray*}
This equation simply tells us that the contribution of the annihilation to the relic density becomes very small for very heavy charged singlet scalar and one needs to have large $Y_{\ell N}$ to produce the correct relic density. On the other hand, for large values of the mass splitting and heavy charged singlet scalar one cannot reproduce the correct relic abundance if one demands perturbativity of the couplings. 
The co-annihilations are more involved in this model as we can have additional contributions that have different dependence on the model parameters. There are two generic co-annihilation channels: $N_R H^\pm \to {\rm SM}$ and $H^\pm H^\mp \to {\rm SM}$. Below, we list the individual contributions and the overall dependence of the corresponding cross section 
\begin{widetext}
\begin{eqnarray*}
N_R H^\pm \to \ell_\alpha^\pm H_{\rm SM} &:& \quad \sigma \propto \lambda_3^2 Y_{\ell_\alpha N}^2, \\
N_R H^\pm \to \ell_\alpha^\pm Z, \ell_\alpha^\pm \gamma, \nu W^\pm &:& \quad \sigma \propto Y_{\ell_\alpha N}^2, \\
H^\pm H^\mp \to \ell_\alpha^\pm \ell_\beta^\mp &:& \quad \sigma \propto |Y_{\ell_\alpha N} Y_{\ell_\beta N}|^2 \mathcal{A}_1 + |Y_{\ell_\alpha N} Y_{\ell_\beta N}| \mathcal{A}_2 + \mathcal{A}_3, \\
H^\pm H^\mp \to q\bar{q} &:& \quad \sigma \propto \lambda_3^2 \mathcal{B}_1 + \lambda_3 \mathcal{B}_2 + \mathcal{B}_3, \\
H^\pm H^\mp \to ZZ, H_{\rm SM} Z, W^\pm W^\mp &:& \quad \sigma \propto \lambda_3^2 \mathcal{C}_1 + \lambda_3 \mathcal{C}_2 + \mathcal{C}_3, \\
H^\pm H^\mp \to H_{\rm SM} H_{\rm SM} &:& \quad \sigma \propto \lambda_3^4 \mathcal{D}_1 + \lambda_3^2 \mathcal{D}_2,
\end{eqnarray*}
\end{widetext}
with $\mathcal{A}_i, \mathcal{B}_i, \mathcal{C}_i$ and $\mathcal{D}_i$ are real-valued coefficients that depend on the dark matter mass, the charged singlet scalar mass and the final-state particles. The co-annihilation becomes very active for quite large $\lambda_3$ and $Y_{\ell N}$ and may even drive the relic density to very small values ($\sim$ orders of magnitudes smaller than the observed abundance). In general, the co-annihilation is dominated by contributions of the following two processes $H^\pm H^\mp \to 2 H_{\rm SM}$ and $N_R H^\pm \to \ell_\alpha^\pm H_{\rm SM}$. In the presence of co-annihilations, the Boltzmann equations become
\begin{widetext}
\begin{eqnarray}
\frac{{\rm d}n_{N_R}}{{\rm d}t} + 3 H n_{N_R} &=& -2  \langle \sigma_{\rm eff} v_r \rangle \bigg[(n_{N_R})^2 - (n_{N_R}^{\rm eq})^2\bigg] + N \Gamma_{H^\pm} n_{H^\pm}, \\
\frac{{\rm d}n_{H^\pm}}{{\rm d}t} + 3 H n_{H^\pm} &=& - \Gamma_{H^\pm} n_{H^\pm},
\label{eq:boltzmann:full}
\end{eqnarray}
\end{widetext}
where $N$ is the mean number of $N_R$ particles, $n_{H^\pm}$ is the number density of $H^\pm$ and $\Gamma_{H^\pm}$ is its total width. Note that here we have replaced the thermally-averaged annihilation cross section in equation \eqref{eq:boltzmann:ann} by the effective cross section 
\begin{eqnarray}
\langle \sigma_{\rm eff} v_r \rangle = \sum_{i,j \in \{N_R, H^\pm\}} \langle \sigma(ij \to {\rm SM}) v_r \rangle \frac{n_i^{\rm eq} n_j^{\rm eq}}{(n_{N_R}^{\rm eq})^2}.
\end{eqnarray}
The relic density of $N_R$ is obtained from the numerical solutions of the coupled Boltzmann equations \eqref{eq:boltzmann:full}. \textsc{MadDM} version 3.0 is used to solve the Boltzmann equations
and compute the relic density of $N_R$ \cite{Ambrogi:2018jqj}. In figure \ref{fig:relicdensity}, we show the values of the coupling $Y_{\ell N}$ consistent with the measurement of the relic density by the \textsc{Planck} collaboration projected on the mass of the dark matter and the mass of the charged singlet scalar. We can see that the relic abundance of the $N_R$ is consistent with the \textsc{Planck} measurement only for very specific regions. If the mass splitting between $H^\pm$ and $N_R$ is large, we need large values of the $Y_{\ell N}$. However, even for $Y_{\ell N}$ near the perturbativity bound the mass splitting can not be arbitrary large: $\Delta_{\rm max} \approx 600~(2000)~{\rm GeV}$ for $M_{N_R} = 10~(100)~{\rm GeV}$.  The relic density becomes almost independent of $Y_{\ell N}$ for large $M_{N_R}$ in the co-annihilation regions. We conclude this section by noting that the model can not reproduce the correct relic density with the standard freeze-out mechanism for the region marked in blue in figure \eqref{fig:relicdensity} as it breaks the perturbativity of the coupling $Y_{\ell N}$.  

\subsection{Direct detection}
\label{sec:direct}
We turn now into a discussion of the constraints from direct detection experiments on the model parameter space. In this model, 
the scattering cross section of $N_R$ off the nucleus with atomic number ($A$) occurs at the one-loop order where the SM Higgs boson plays the role 
of a portal.
The generic formula for the spin-independent cross section is given by\footnote{The spin-dependent cross section is very small in our model as the exchanged particle is the SM Higgs boson which is a scalar particle with $J^P=0^+$. Nevertheless, we will compute this observable for some benchmark scenarios and estimate their consistency with the current bounds from the \texttt{PICO} experiment \cite{PICO:2017tgi}.}
\begin{eqnarray}
 \sigma_{\rm SI} = \frac{4}{\pi} \mu_{A}^2 \bigg( Z \cdot \mathcal{S}_p + (A - Z) \cdot \mathcal{S}_n \bigg)^2,
\end{eqnarray}
with $\mathcal{S}_{p,n}$ being the scalar current nucleon ($p/n$) form factors and $\mu_{A} \equiv M_{N_R} m_{A}/(M_{N_R} + m_{A})$ is 
the reduced mass of the $N_R$--$A$ system. The nucleon form factors have two contributions: {\it (i)} from particle physics which is connected to
the scattering amplitude of the $N_R$--$(q/g)$ process and {\it (ii)} from low-energy nuclear physics 
that are computed using chiral perturbation theory  \cite{DelNobile:2013sia, Hill:2014yxa, Bishara:2017pfq, Ellis:2018dmb}. 

\begin{figure*}
    \centering
    \includegraphics[width=0.56\linewidth]{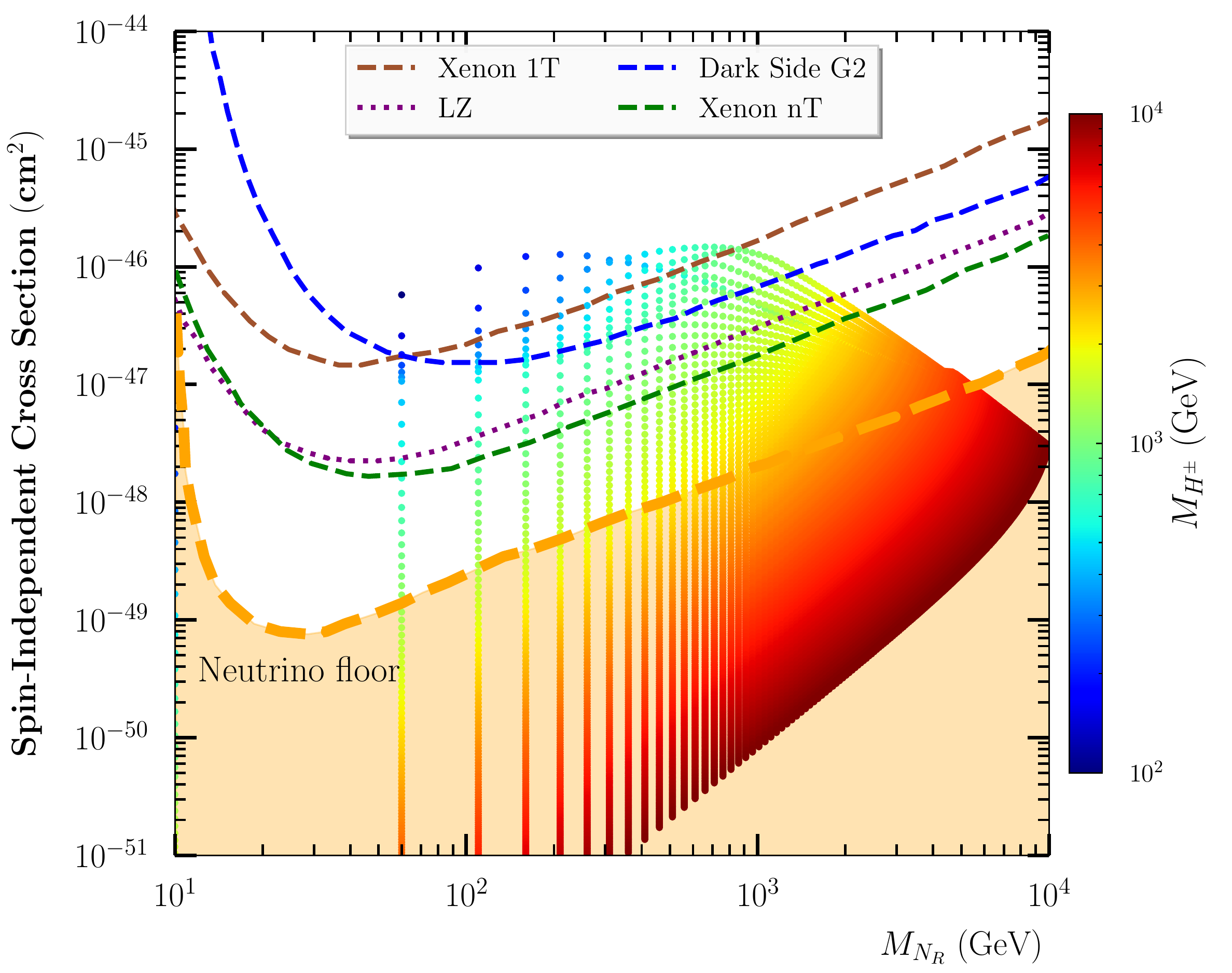}
    \caption{The spin-independent cross section as function of the dark matter mass $M_{N_R}$ while the colored scatter points correspond to the charged singlet scalar mass $M_{H^\pm}$. In the same plot, we show the current bounds from \texttt{Xenon} 1T \cite{Aprile:2018dbl} in dashed sienna, and the future expectations from \texttt{Xenon} nT \cite{XENON:2020kmp}, LUX LZ \cite{LZ:2018qzl} and DarkSide G2 \cite{DarkSide-20k:2017zyg}. The shaded orange area marked by `neutrino floor' corresponds to the backgrounds from the coherent scattering with solar neutrinos, atmospheric neutrinos and supernova neutrinos \cite{Billard:2013qya}. The spin-independent cross section was scaled by a factor of 
    $\xi_{\rm Planck} = \Omega_{N_R} h^2/\Omega_{\rm Planck} h^2$ with $\Omega_{\rm Planck} h^2 \approx 0.12$. All the calculations were performed for $Y_{\ell N} = 2$ and $\lambda_3 = 4$.}
    \label{fig:SI:bounds}
\end{figure*}

The generic formula of $f_{p,n}$ is given by 
\begin{eqnarray}
 \mathcal{S}_{p,n} = m_{p,n} \sum_{u,d,s} \frac{\mathcal{A}_q}{m_q} \mathcal{S}^{q}_{p,n} + \frac{2}{27} m_{p,n} \mathcal{S}_{p,n}^{g} \sum_{c,b,t} \frac{\mathcal{A}_q}{m_q},
\end{eqnarray}
where $m_p = 938.27~{\rm MeV}$, $m_n = 939.56~{\rm MeV}$ are the proton and neutron masses respectively. The values of the scalar nucleon low energy form factors are chosen to be \cite{Backovic:2015cra}
\begin{widetext}
\begin{eqnarray*}
\mathcal{S}_p^{u} = 1.53 \times 10^{-2}, \quad \mathcal{S}_p^{d} = 1.91 \times 10^{-2}, \quad  \mathcal{S}_p^{s} = 4.47\times 10^{-2}, \quad \mathcal{S}_p^{g} = 1 - \sum_{q=u,d,s} \mathcal{S}_p^q = 92.09 \times 10^{-2}, \\
\mathcal{S}_n^{u} = 1.10 \times 10^{-2}, \quad \mathcal{S}_n^{d} = 2.73 \times 10^{-2}, \quad  \mathcal{S}_n^{s} = 4.47\times 10^{-2}, \quad \mathcal{S}_n^{g} = 1 - \sum_{q=u,d,s} \mathcal{S}_n^q = 91.70 \times 10^{-2}.
\end{eqnarray*}
\end{widetext}
The parton-level scattering amplitude is
\begin{eqnarray}
 \mathcal{M}_{q N_R \to q N_R} = \mathcal{A}_q \bar{\psi}_q(p_{\rm out}) \psi_q(p_{\rm in}),
\end{eqnarray}
where $\mathcal{A}_q$ is connected to the non-hadronic part of the amplitude. The term
$\bar{\psi}_q(p_{\rm out}) \psi_q(p_{\rm in})$ should be incorporated in a hadronic current $\langle \mathcal{N} | \cdot | \mathcal{N} \rangle$
\begin{eqnarray}
    \langle \mathcal{N} | \bar{\psi}_q \psi_q | \mathcal{N} \rangle = \Bigg\{\begin{array}{lr}
        \frac{m_\mathcal{N}}{m_q} \cdot \mathcal{S}_{\mathcal{N}}^{q}, & \text{for } q=u,d,s, \\
        \frac{2}{27} \frac{m_\mathcal{N}}{m_q} \cdot \mathcal{S}_{\mathcal{N}}^{g}, & \text{for } q=c,b,t,
        \end{array}
\end{eqnarray}
where $\mathcal{N} = p,n$. The model-dependent non-hadronic form factor is given by
\begin{eqnarray}
 \mathcal{A}_q = \frac{\tilde{y}(Q^2\approx0)}{M_{H_{\rm SM}}^2} \cdot \frac{m_q}{v} \bar{\psi}_{N_R}(k_{\rm out}) \psi_{N_R}(k_{\rm in}),
\end{eqnarray}
here $\tilde{y}(Q^2\approx 0)$ is the effective $H_{\rm SM} N_R N_R$ coupling computed in the low energy limit. With the help of the 
Package X \cite{Patel:2015tea}, we can obtain it from equation \eqref{eq:yHNN} 
\begin{widetext}
\begin{eqnarray}
\tilde{y}(Q^2 \approx 0) \simeq -\frac{\lambda_3 v |Y_{\ell N}|^2}{16 \pi M_{H^\pm}} \frac{1}{\varrho_N} \bigg[1 - \left(1 - \varrho_N^{-2} \right) \log{\left(1 - \varrho_N^2 \right)} \bigg] \equiv -\frac{\lambda_3 v |Y_{\ell N}|^2}{16 \pi M_{H^\pm}} \mathcal{H}(\varrho_N), 
\label{eq:effective:HNN}
\end{eqnarray}
\end{widetext}
where $\varrho_N = M_{N_R}/M_{H^\pm}$. $\mathcal{H}(x)$ is monotonous and increasing function of $x$ in the interval $[0,1]$ and has the following limits $\lim_{x\to 0} \mathcal{H}(x) = 0$ and $\lim_{x\to 1} \mathcal{H}(x) = 1$. Note that the first limit correspond to a small dark matter mass and a heavy charged scalar for which the model cannot reproduce the correct relic abundance while the second limit corresponds to the nearly degenerate scenario where co-annihilation is the most active component in the relic abundance calculation. In addition the effective coupling involves an extra suppression by $1/M_{H^\pm}$ which simply means that the direct detection spin-independent cross section is always below the neutrino floor for heavy $H^\pm$. From equation \eqref{eq:effective:HNN} one also expect that the spin-independent cross section is always proportional to $|Y_{\ell N}|^4$. Therefore, large $Y_{\ell N}$ regions with large $\sigma_{\rm SI}$ would also correspond to small relic density (which is proportional to $1/|Y_{\ell N}|^4$)\footnote{This is consistent with our previous finding in \cite{Jueid:2020yfj} where a strong anti-correlation between $\sigma_{\rm SI}$ and $\Omega_{N_R} h^2$ was observed.} and for these scenarios $\sigma_{\rm SI}$ needs to be scaled by a factor $\xi_{\rm Planck} \equiv \Omega_{N_R} h^2/\Omega_{\rm Planck} h^2$. This means that the spin-independent cross section would always be consistent with the current \texttt{Xenon} 1T bounds \cite{Aprile:2018dbl} for most regions of the parameter space as we can see clearly in figure \ref{fig:SI:bounds}.  

\begin{table*}[hbt!]
\setlength\tabcolsep{12pt}
\begin{center}
\begin{adjustbox}{max width=0.95\textwidth}
\begin{tabular}{lcccc}
\toprule
\multicolumn{1}{c} { Benchmark point } & BP1 & BP2 & BP3 & BP4 \\
\midrule
\multicolumn{5}{c}{\textit{Parameters}} \\
\midrule
$M_{N_R}~({\rm GeV})$    & $50$  &   $200$ & $598$ & $1000$ \\
$M_{H^\pm}~({\rm GeV})$ & $500$ & $500$ & $600$  & $1500$  \\
$Y_{Ne}$       & $10^{-4}$ &     $5 \times 10^{-4}$ & $10^{-3}$ & $5 \times 10^{-3}$ \\
$Y_{N\mu}$   & $2.8$      &     $1.6$  & $1$ & $2$  \\
$Y_{N\tau}$   & $5 \times 10^{-2}$  & $5 \times 10^{-1}$ & $5 \times 10^{-1}$ & $2$  \\
$\lambda_3$  & $4$         &      $5$ & $5$ & $6$  \\
\midrule
\multicolumn{5}{c}{{\it Decays of $H^\pm$}} \\
\midrule
${\rm BR}(H^\pm \to e N_R)$     & $1.27 \times 10^{-9}$ & $8.89 \times 10^{-8}$ & $8.98 \times 10^{-7}$ & $3.12 \times 10^{-6}$  \\
${\rm BR}(H^\pm \to \mu N_R)$   & $99.96 \times 10^{-2}$ & $91.10 \times 10^{-2}$ & $89.70 \times 10^{-2}$ & $50.0 \times 10^{-2}$  \\
${\rm BR}(H^\pm \to \tau N_R)$  & $3.18 \times 10^{-4}$ & $8.89 \times 10^{-2}$ & $10.29 \times 10^{-2}$ & $49.99 \times 10^{-2}$  \\
$\Gamma_{H^\pm}~({\rm GeV})$ & $76.45$ & $19.72$ & $5.88 \times 10^{-4}$ & $73.68$ \\
$\Gamma_{H^\pm}/M_{H^\pm}$ & $15.29 \times 10^{-2}$ & $3.94 \times 10^{-2}$ & $9.81 \times 10^{-7}$ & $4.91 \times 10^{-2}$  \\
\midrule
\multicolumn{5}{c}{${\rm BR}(\ell_\alpha \to \ell_\beta \gamma)$ and ${\rm BR}(\ell_\alpha \to 3 \ell_\beta)$} \\
\midrule
${\rm BR}(\mu \to e \gamma)$    & $2.68 \times 10^{-14}$ & $1.51 \times 10^{-13}$ & $4.31 \times 10^{-14}$ & $1.89 \times 10^{-13}$ \\
${\rm BR}(\tau \to e \gamma)$   & $1.52 \times 10^{-18}$ & $2.64 \times 10^{-15}$ & $1.92 \times 10^{-15}$ & $3.38 \times 10^{-14}$ \\
${\rm BR}(\tau \to \mu \gamma)$ & $1.17 \times 10^{-9}$ & $2.64 \times 10^{-8}$ & $1.87 \times 10^{-9}$ & $5.28 \times 10^{-9}$  \\
\midrule
${\rm BR}(\mu \to eee)$ & $1.47 \times 10^{-16}$ & $8.21 \times 10^{-16}$ & $2.27 \times 10^{-16}$ &  $1.01 \times 10^{-15}$  \\
${\rm BR}(\tau \to eee)$ & $1.51 \times 10^{-20}$ &  $2.58 \times 10^{-17}$ & $1.85 \times 10^{-17}$ & $3.29 \times 10^{-16}$  \\
${\rm BR}(\tau \to \mu\mu\mu)$ & $1.21 \times 10^{-8}$ & $9.79 \times 10^{-9}$ & $2.63 \times 10^{-12}$ & $1.17 \times 10^{-9}$  \\
\midrule
\multicolumn{5}{c}{${\rm BR}(H_{\rm SM} \to \ell_\alpha \ell_\beta)$} \\
\midrule
${\rm BR}(H_{\rm SM} \to \mu\tau)$ & $2.31 \times 10^{-8}$ & $1.18 \times 10^{-6}$ & $2.22 \times 10^{-7}$ & $5.24 \times 10^{-7}$ \\
${\rm BR}(H_{\rm SM} \to e\tau)$ & $2.95 \times 10^{-17}$ & $1.15 \times 10^{-13}$ & $2.22 \times 10^{-13}$ & $3.27 \times 10^{-12}$ \\
\midrule
\multicolumn{5}{c}{{\it Dark matter observables}} \\
\midrule
$\Omega_{N_R} h^2$      & $9.84 \times 10^{-2}$ & $9.25 \times 10^{-2}$ & $2.11 \times 10^{-3}$ & $8.53 \times 10^{-2}$  \\
$\langle \sigma \upsilon \rangle~({\rm cm}^2)$ & $2.40 \times 10^{-9}$ & $2.55 \times 10^{-9}$ & $7.32 \times 10^{-8}$ & $2.69 \times 10^{-9}$   \\
$\sigma_{\rm SI}^p~({\rm cm}^2)$ & $1.60 \times 10^{-47}$ & $3.45 \times 10^{-47}$ & $2.28 \times 10^{-48}$  & $1.47 \times 10^{-46}$  \\
$\sigma_{\rm SD}^p~({\rm cm}^2)$       & $6.51 \times 10^{-62}$ & $6.29 \times 10^{-62}$ & $1.98 \times 10^{-65}$ & $8.29 \times 10^{-60}$  \\
$\texttt{XENON1T}$      & $\checkmark$  & $\checkmark$ & $\checkmark$ & $\checkmark$  \\
PICO                          & $\checkmark$ & $\checkmark$ & $\checkmark$ & $\checkmark$  \\
DarkSide~G2           & $\checkmark$ & X & $\checkmark$  & X \\
$\texttt{LZ}$		              &  X &  X & $\checkmark$ & X  \\
$\texttt{Neutrino~floor}$          & X &  X & X & X  \\
\bottomrule
\end{tabular}
\hspace{0.2cm}
\end{adjustbox}
\end{center}
\caption{Characteristics of the four benchmark points in our model. Here, we show the values of the independent parameters, the decay branching ratios and total width of the charged singlet scalar, the CLFV decay branching ratios and dark-matter observables. A checkmark ($\checkmark$) indicate that the parameter point yields a smaller $\sigma_{\rm SI}$ than the experimental bound (present or expected) while a cross mark (X) indicates that $\sigma_{\rm SI}$ is above the experimental bound.}
\label{tab:BSs}
\end{table*} 

\subsection{Higgs invisible decay}
\label{sec:invisible}
The Higgs invisible decay occurs at the one-loop order with the exchange of charged scalar and right-handed fermion. The partial decay width is given by 
\begin{eqnarray}
\Gamma(H_{\rm SM} \to N_R N_R) = \frac{M_{H_{\rm SM}} |\tilde{y}|^2}{8\pi} \bigg(1 - \frac{4 M_{N_R}^2}{M_{H_{\rm SM}}^2} \bigg)^{3/2}, \hspace{0.5cm}
\label{eq:width:Hinv}
\end{eqnarray}
with $\tilde{y}$ is the one-loop induced effective $H_{\rm SM}$--$N_R$--$N_R$ coupling which is given by
\begin{eqnarray}
\tilde{y} = \frac{\lambda_3 \upsilon M_{N_R}}{16 \pi^2} \sum_\ell |Y_{\ell N}|^2 (C_0 + C_2),
\label{eq:yHNN}
\end{eqnarray}
with $C_i \equiv C_i(M_{N_R}^2,M_{H_{\rm SM}}^2,M_{N_R}^2,m_\ell^2,M_{H^\pm}^2,M_{H^\pm}^2), i=0,2$ being the Passarino-Veltman three-point functions \cite{Passarino:1978jh}. The computation of the Feynman amplitudes has been performed using \textsc{FeynArts}, \textsc{FormCalc}, and \textsc{LoopTools} \cite{Hahn:1998yk, Hahn:2000kx}. We have used a \textsc{Python} interface to \textsc{LoopTools} to evaluate numerically the one-loop integrals\footnote{\texttt{pylooptools} is a \textsc{Python} binding to \textsc{LoopTools} and can be found in this \texttt{github} directory: \texttt{https://github.com/djukanovic/pylooptools.git}.}. We define the Higgs invisible branching ratio as 
\begin{eqnarray}
B_{\rm inv} \equiv \frac{\Gamma(H_{\rm SM} \to N_R N_R)}{\Gamma(H_{\rm SM} \to N_R N_R) + \Gamma_H^{\rm SM}},
\label{eq:BRinv}
\end{eqnarray}
where $\Gamma_H^{\rm SM} = 4.07~{\rm MeV}$. Using equations \eqref{eq:width:Hinv}, \eqref{eq:yHNN} and \eqref{eq:BRinv}, we can obtain bounds on the coupling $Y_{\ell N}$. The bound is analytically defined by
\begin{eqnarray*}
Y_{\ell N} < \bigg(\frac{2048 \pi^5 \Gamma_{H}^{\mathrm{SM}}}{\beta_N^{3/2} M_{H_{\rm SM}} \lambda_3^2 \upsilon^2 M_{N_R}^2 |C_0 + C_2|^2 \left(\frac{1}{B_{\rm bound}} - 1\right)}\bigg)^{1/4},
\end{eqnarray*}
where $\beta_N \equiv (1 - 4 M_{N_R}^2/M_{H_{\rm SM}}^2)$ and $B_{\rm bound}$ is the upper bound on ${\rm BR}_{\rm inv}$. \\
\begin{figure*}[!t]
    \centering
    \includegraphics[width=0.62\linewidth]{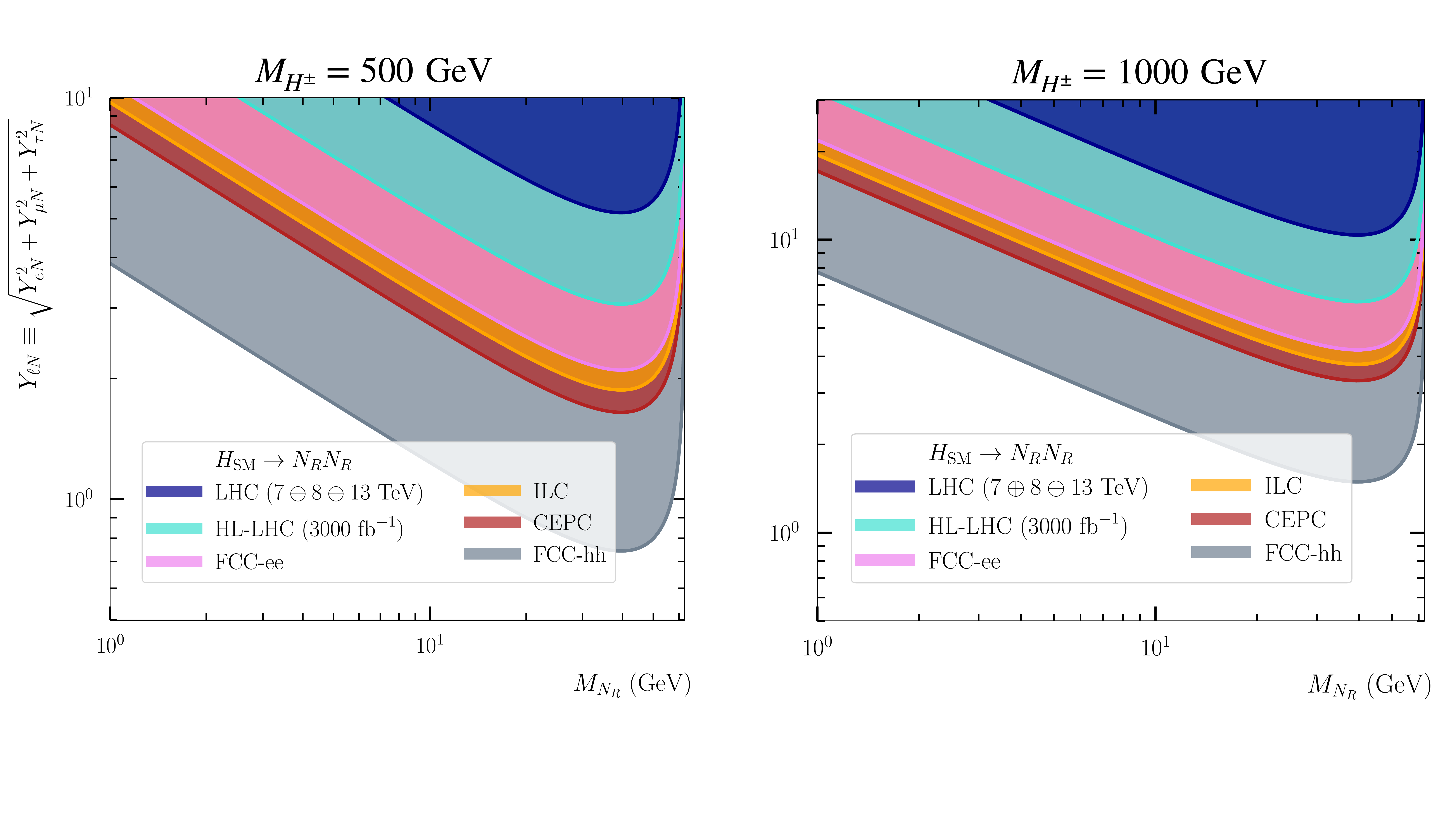}
    \vspace{-0.9cm}
    \caption{The present and future exclusions on values of $Y_{\ell N}$ for $M_{H^\pm} = 500~{\rm GeV}$ and $\lambda_3 = 4$. Here we show the contours obtained from the LHC (navy), HL-LHC (turquoise), FCC--ee (magenta), ILC (orange), CEPC (dark red) and FCC--hh (gray). All the bounds were obtained assuming SM Higgs boson mass of $M_{H_{\rm SM}} = 125~{\rm GeV}$, and SM Higgs boson production rates. The Higgs diphoton rate is assumed to be equal to the SM prediction at LO.}
    \label{fig:Higgs:invisible}
\end{figure*}

Searches for Higgs invisible decays have been carried by the ATLAS and the CMS collaborations \cite{ATLAS:2015ciy, CMS:2016dhk, Sirunyan:2018owy}. The strongest and up-to-date stringent bound on $\textrm{B}_{\mathrm{inv}}$ was reported by the \textsc{Cms} collaboration using  a combination of  previous Higgs to invisible decay searches at $7, 8$ and $13$ TeV, where it has been found $B_{\mathrm{inv}} < B_{\rm bound} = 0.19$ at $95\%$ CL \cite{Sirunyan:2018owy} assuming that the rates of the Higgs boson production are equal to the SM predictions. On the other hand, several groups have carried global analyses using recent Higgs boson measurements and obtained stringent limits \cite{Belanger:2013xza, Ellis:2013lra}. Finally, several studies have been devoted to the projected sensitivities of the future collider experiments to Higgs invisible decays from HL--LHC \cite{CMS-PAS-FTR-16-002}, FCC--ee \cite{Cerri:2016bew}, ILC \cite{Asner:2013psa}, CEPC \cite{CEPCStudyGroup:2018ghi} and FCC--hh \cite{Selvaggi:2018obt}. In figure \ref{fig:Higgs:invisible}, we show the excluded values of $Y_{\ell N}$ from present and future bounds on ${\rm BR}_{\rm inv}$ assuming $M_{H^\pm} = 500~{\rm GeV}$ and $\lambda_3 = 4$\footnote{This choice of the charged singlet scalar mass is consistent with the limits from searches of sleptons at the LHC. We note that increasing $M_{H^\pm}$ would weaken the bounds on $Y_{\ell N}$ from Higgs invisible decays.}. As we can see clearly that the present bounds are extremely weak which excludes $Y_{\ell N} \sim 6$ for $M_{N_R} \sim 49~{\rm GeV}$. The future experiments are expected to exclude smaller values of $Y_{\ell N}$; {\it e.g.} FCC--hh can exclude values up to $0.7$ for $M_{N_R} \sim 49~{\rm GeV}$.

\subsection{Benchmark points}
\label{BPs}
From the discussions in sections \ref{sec:model} and \ref{sec:DM}, we can conclude the following:
\begin{itemize}
    \item The scalar singlet cannot be lighter than $440$ GeV for mass splittings with the dark matter of order $\leq 80$ GeV.
    \item CLFV can constrain only the product of the Yukawa-type couplings and not their individual values. Therefore, benchmark points have to be chosen.
    \item DM direct detection constraints are not very strong as expected since the spin-independent cross section is one-loop induced. 
    \item The constraints from the consistency with the measurement of the DM relic density forbids large mass splittings if the Yukawa-type couplings are of order $\mathcal{O}(1)$. 
\end{itemize}

The benchmark points used in the discussion of the general features of DM production at Muon colliders are shown in Table \ref{tab:BSs}. There are four of these benchmarks and each one has its own phenomenological implications. 

\paragraph{BP1.} This benchmark point is characterised by a relatively light DM ($M_{N_R} = 50~{\rm GeV}$) and a charged singlet mass in near the exclusion limit reported on by the LHC (see figure \ref{fig:collider:bounds}). On the other hand, the Yukawa-type couplings are chosen such that $Y_{\mu N} \gg Y_{\tau N} > Y_{e N}$. This choice leads to a charged singlet decaying predominantly into $\mu N_R$ with a branching fraction approaching $100\%$. On the other hand, the charged lepton flavour violating decays of charged leptons are such that ${\rm BR}(\tau \to e \gamma)$ is well below the sensitivity reach of future experiments in the foreseeable future. The other branching ratios, are below the current experimental bounds but can be tested in the near future. For DM observables, the relic density for this BP is about $90\%$ of the observed abundance and the spin-independent DM-nucleon cross section is below \texttt{Xenon1T} bound and the expected DarkSide G2 bound but can be excluded or discovered by \texttt{LZ}. 

\paragraph{BP2.} For this point, we choose $M_{H^\pm} = 500~{\rm GeV}$ and $M_{N_R} = 200$ GeV. The Yukawa-type couplings are chosen using the same hierarchy as BP1 but with relatively different values, {\it i.e.}, $Y_{\mu N} = 1.6$, $Y_{\tau N} = 5 \times 10^{-1}$ and $Y_{e N} = 5 \times 10^{-4}$. This leads to the following branching ratios ${\rm BR}(H^\pm \to \mu^\pm N_R) \simeq 91\%$, ${\rm BR}(H^\pm \to \tau^\pm N_R) \simeq 9\%$ and ${\rm BR}(H^\pm \to e^\pm N_R) \simeq 0\%$. The charged singlet is narrow in this case as $\Gamma_{H^\pm}/M_{H^\pm} \simeq 0.04$. The CLFV decays of charged leptons exhibits similar features as in BP1 with the exception that ${\rm BRs}$ of $\tau \to \mu N_R$ and $\mu \to e \gamma$ can be probed in the future experiments as they are slightly below the current bounds. The spin-independent DM-nucleon cross section can be tested by the DarkSide G2 experiment.

\paragraph{BP3.} For this point, we choose the following values of the particle masses: $M_{H^\pm} = 600$ GeV and $M_{N_R} = 598$ GeV and therefore a small mass splitting of $2$ GeV. We choose the following values for the Yukawa-type couplings: $\{Y_{\mu N}, Y_{\tau N}, Y_{e N}\} = \{1, 0.5, 10^{-3}\}$ which leads to the following branching fractions: ${\rm BR}(H^\pm \to \mu^\pm N_R) \simeq 90\%$, ${\rm BR}(H^\pm \to \tau^\pm N_R) \simeq 10\%$ and ${\rm BR}(H^\pm \to e^\pm N_R) \simeq 0\%$. On the other hand, the branching ratios of CLFV decays of $\mu \to e\gamma$ and $\tau \to \mu \gamma$ can be tested in future experiments. Since the mass splitting is equal to $2$ GeV, the most active component in the calculation of the relic density comes from coannihilation-based freezout and therefore the choice of $\lambda_3$ is pivotal in this case. We found that for this BP, the relic density of the $N_R$ is below $2\%$ of the total observed DM relic density. Finally, this BP is not sensitive to the direct detection experiments and the cross section is above the neutrino floor.  

\paragraph{BP4.} Here, we choose relatively heavy DM and charged singlet scalar; $M_{N_R} = 1000$ GeV and $M_{H^\pm} = 1500$ GeV. The Yukawa-type couplings are chosen such that $Y_{\mu N} = Y_{\tau N} = 2 \gg Y_{e N} = 5 \times 10^{-3}$. With this choice, one gets: ${\rm BR}(H^\pm \to \mu^\pm N_R) \simeq {\rm BR}(H^\pm \to \tau^\pm N_R) \simeq 50\%$ while ${\rm BR}(H^\pm \to e^\pm N_R)$. Similar features to BP1 and BP2 are observed for CLFV and DM phenomenology.


\section{Production of dark matter at muon colliders}
\label{sec:prod:NR}

\subsection{Total cross sections}

In this section, we discuss the general features of DM production at muon colliders\footnote{The cross sections for both DM and charged scalar production at muon colliders are computed at Leading order using \textsc{Madgraph\_aMC@NLO} \cite{Alwall:2014hca} with a UFO model file \cite{Degrande:2011ua} that can be found in the FeynRules model database \href{FR}{https://feynrules.irmp.ucl.ac.be/wiki/ModelDatabaseMainPage}.}. In this model, DM can be produced through a variety of processes:
\begin{itemize} 
\item DM production in association with one SM particle dubbed as mono-X. Given the nature of the interaction Lagrangian and the fact that the initial state has a zero total electric charge, DM can only produced in association with one neutral boson. Therefore, we have mono-$\gamma$, mono-$Z$ and mono-Higgs (a full analysis of these channels will be done in future work \cite{Jueid:2023txs}).

\item DM production in association with two SM particles. For this category, we have seven different processes. The rates of those processes are slightly smaller than the mono-X production channels. However, these processes have smaller backgrounds (a full analysis of these channels will be done in future work \cite{Jueid:2023tzy}). 

\item DM production in association with three SM particles. The rates of the $N_R N_R$ in association with three SM particles are even smaller than the other two categories. The signal-to-background optimisation for these channels are even more complicated while the backgrounds, on the other hand, are extremely small.
\end{itemize}

\begin{figure*}[!t]
\centering
\includegraphics[width=0.49\linewidth]{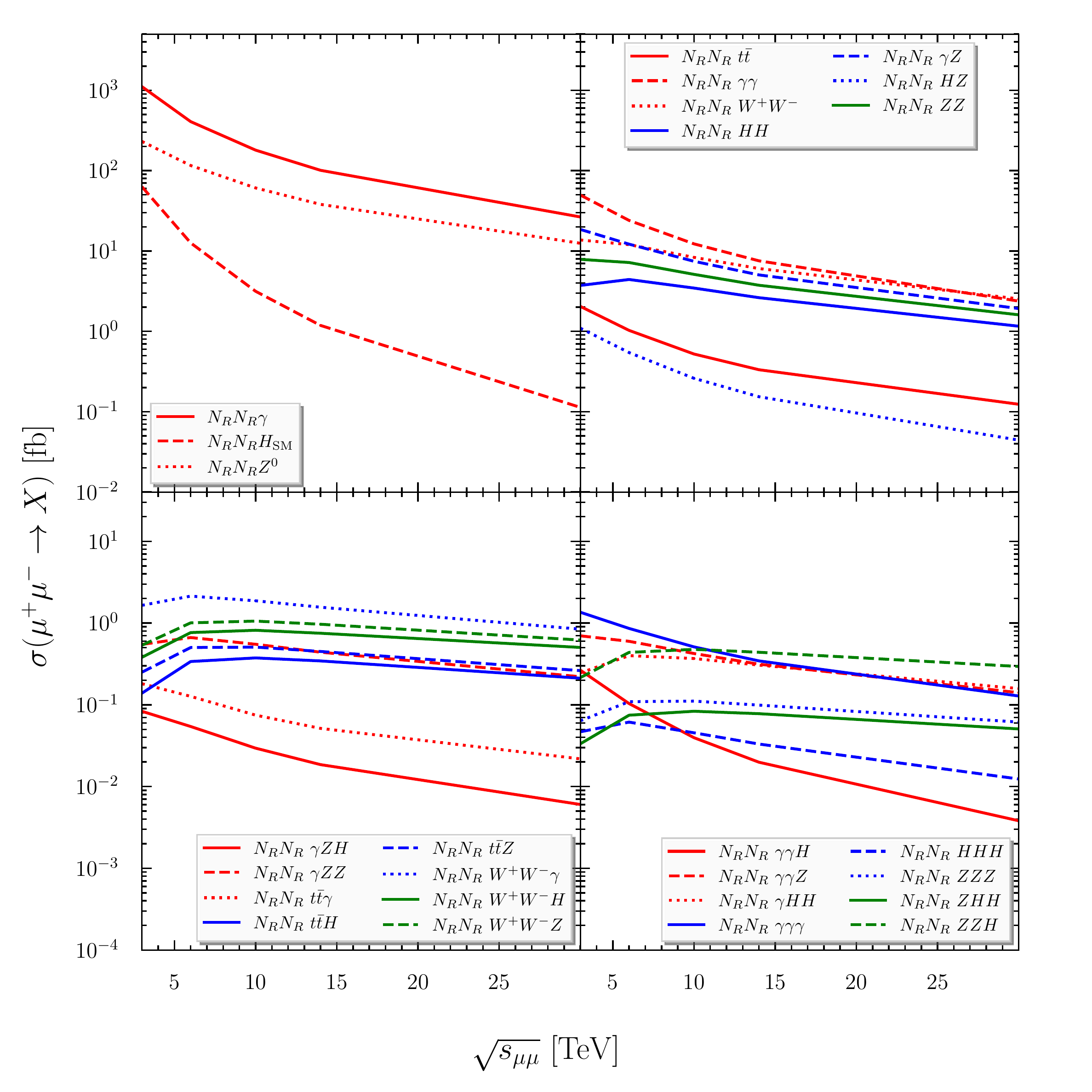}
\hfill
\includegraphics[width=0.49\linewidth]{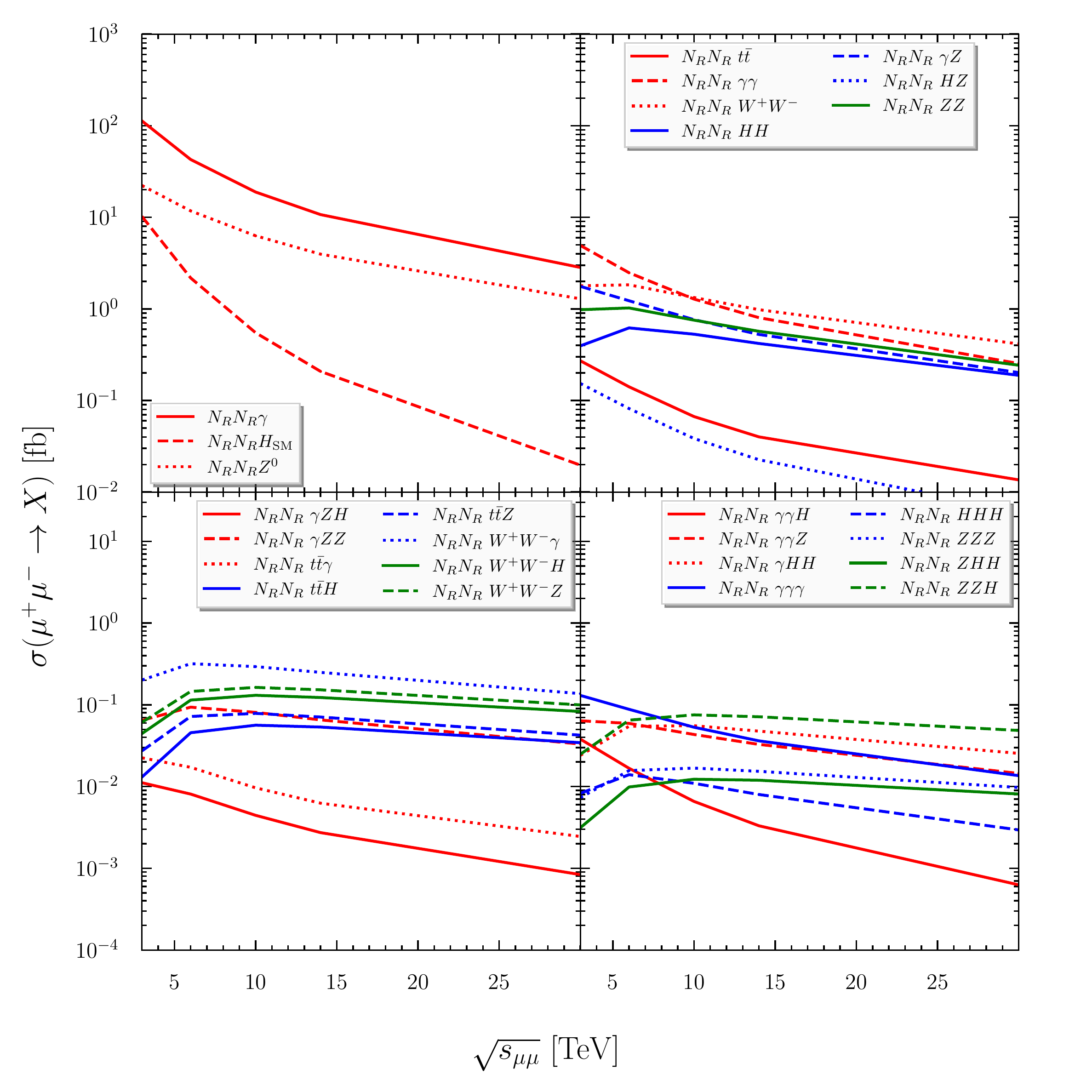}
\vfill
\includegraphics[width=0.49\linewidth]{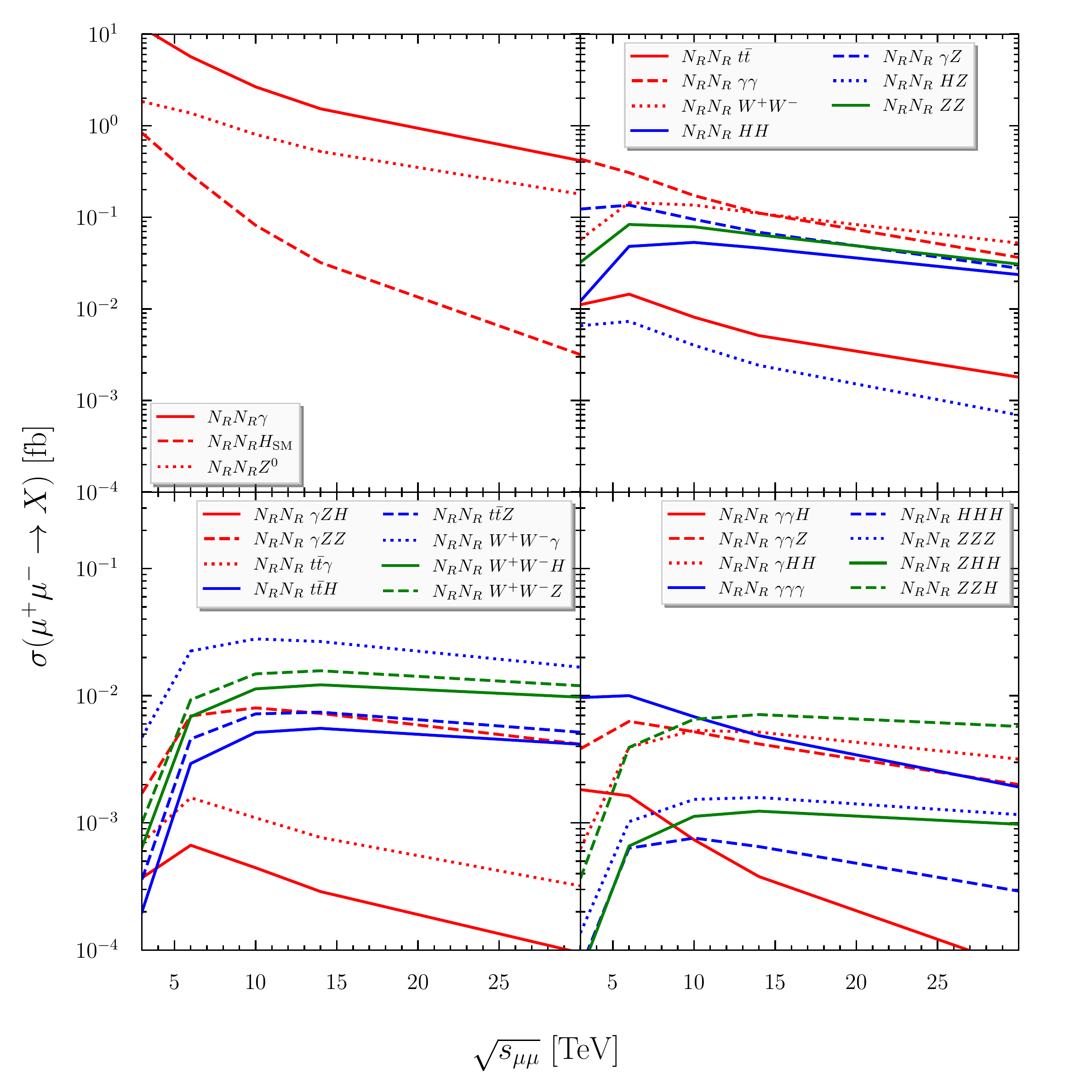}
\hfill
\includegraphics[width=0.49\linewidth]{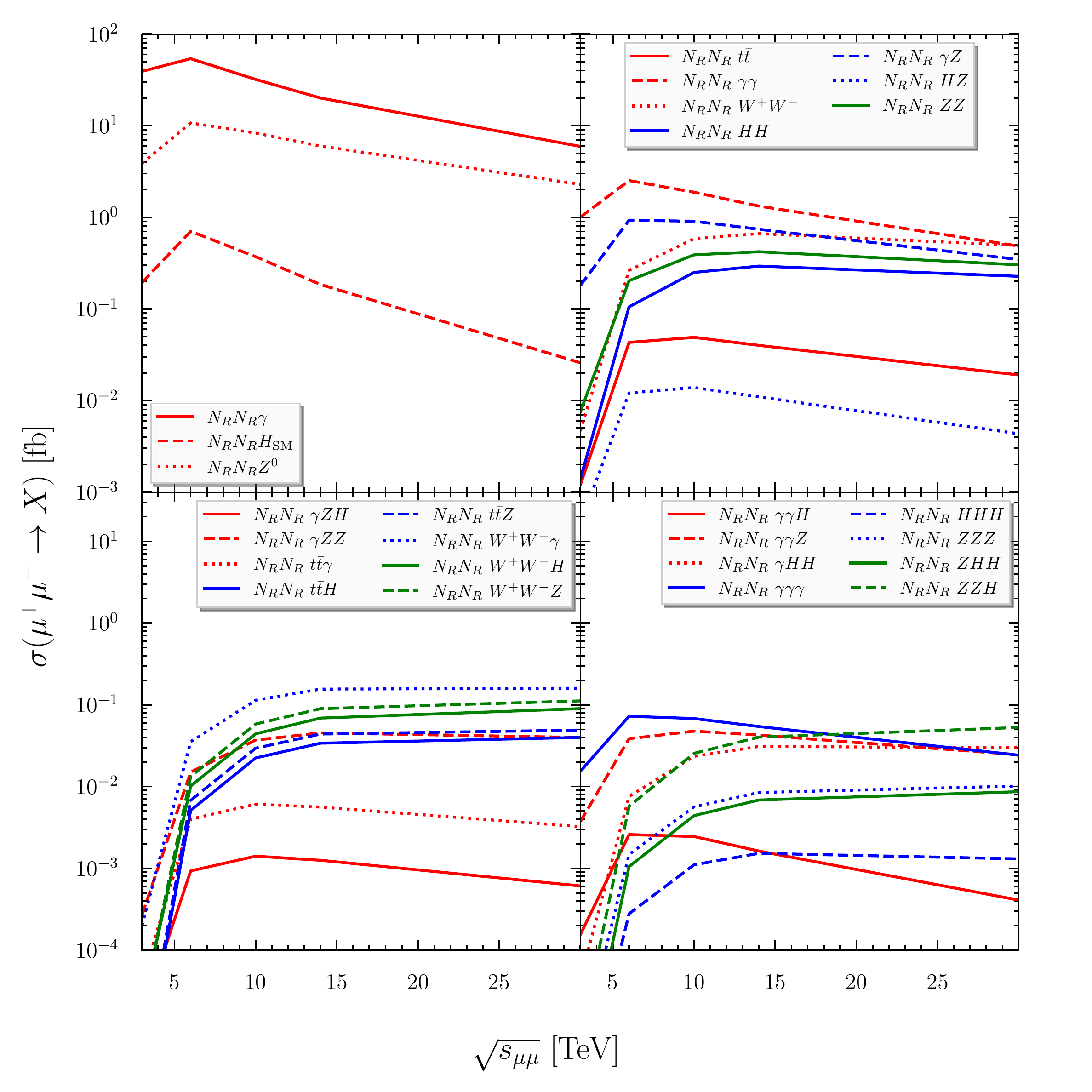}
\caption{Production cross section of $N_R N_R + X$ as a function of the center-of-mass energy ($\sqrt{s_{\mu\mu}}$) for the benchmark points BP1 (left upper panel), BP2 (right upper panel), BP3 (left lower panel) and BP4 (left lower panel). For each pane, we show the production cross section for $N_R N_R$ plus one SM particle, plus two SM particles and in association with three SM particles.}
\label{fig:XS:NR}
\end{figure*} 

\begin{table*}[!t]
\setlength\tabcolsep{8pt}
\begin{center}
\begin{adjustbox}{max width=\textwidth}
\begin{tabular}{l c c c l}
\toprule
\toprule
& & & & \\ [-0.4ex]
 & \multicolumn{3}{c}{$\sigma \times {\rm BR}$~[fb]~({number of events})} & Dominant backgrounds \\ [2.3ex]
\toprule
\toprule
& & & & \\ [-0.4ex]
 $\sqrt{s_{\mu\mu}}$ [TeV] &  $3$ & $10$ & $30$ & \\ [2.3ex]
\toprule
\toprule
& & & & \\ [-0.2ex]
\multirow{4}{*}{$N_R N_R \gamma$} & $1.11 \times 10^{3}~(1.11\times 10^6)$ & $1.80 \times 10^2~(1.80\times 10^6)$ & $2.65 \times 10^1~(2.65\times 10^6)$ & \multirow{4}{*}{$\nu\bar{\nu} + \gamma, 2 \nu\bar{\nu} + \gamma$} \\ 
& $1.13 \times 10^2~(1.13\times 10^5)$ & $1.88 \times 10^1~(1.88\times 10^5)$  & $2.83 \times 10^0~(2.83\times 10^5)$ &  \\
& $1.18 \times 10^1~(1.18\times 10^3)$ & $2.65 \times 10^0~(2.65 \times 10^4)$  & $0.41 \times 10^0~(4.10\times 10^4)$ &  \\ 
& $3.92 \times 10^1~(3.95\times 10^4)$ & $3.20 \times 10^1~(3.20 \times 10^5)$  & $5.94 \times 10^0~(5.94 \times 10^5)$ & \\ [0.2ex]
\midrule
& & & &  \\ [-0.2ex]
\multirow{4}{*}{$N_R N_R Z(\to \ell\ell)$} & $1.68 \times 10^{1}~(1.68\times 10^4)$ & $4.44 \times 10^0~(4.44\times 10^4)$ & $0.91 \times 10^0~(9.10 \times 10^4)$ &  \\ 
& $1.62 \times 10^0~(1.62\times 10^3)$ & $0.46 \times 10^0~(4.58 \times 10^3)$  & $9.39 \times 10^{-2}~(9.39\times 10^3)$ & $\gamma/Z(\to \ell\ell) + \nu\bar{\nu}$  \\
& $0.13 \times 10^0~(0.13 \times 10^3)$ & $0.58 \times 10^{-1}~(0.58\times 10^3)$  & $1.30 \times 10^{-2}~(1.30 \times 10^3)$ & $W (\to \ell\nu_\ell) W (\to \ell\nu_\ell)$  \\ 
& $0.28 \times 10^0~(0.28 \times 10^3)$ & $0.61 \times 10^0~(0.61 \times 10^4)$  & $0.17 \times 10^0~(1.70 \times 10^4)$ & \\ 
& & & & \\ [-0.2ex]
\midrule
& & & &  \\ [-0.2ex]
\multirow{4}{*}{$N_R N_R Z(\to q\bar{q})$} & $1.59 \times 10^{2}~(1.59\times 10^5)$ & $4.20 \times 10^1~(4.20\times 10^5)$ & $8.61 \times 10^0~(8.61 \times 10^5)$ &  \\ 
& $1.53 \times 10^1~(1.53\times 10^4)$ & $4.33 \times 10^0~(4.33 \times 10^4)$  & $0.89 \times 10^{0}~(8.89\times 10^4)$ & $\gamma/Z(\to q\bar{q}) + \nu\bar{\nu}, H_{\rm SM}(\to b\bar{b}) + \nu\bar{\nu}$  \\
& $1.26 \times 10^0~(1.26 \times 10^3)$ & $0.55 \times 10^{0}~(5.54\times 10^3)$  & $0.12 \times 10^{0}~(1.23 \times 10^4)$ & $W (\to \ell\nu_\ell) W (\to q\bar{q}), t\bar{t}$  \\ 
& $2.67 \times 10^0~(2.67 \times 10^3)$ & $5.73 \times 10^0~(5.73 \times 10^4)$  & $1.57 \times 10^0~(1.57 \times 10^5)$ & \\ 
& & & & \\ [-0.2ex]
\midrule
& & & &  \\ [-0.2ex]
\multirow{4}{*}{$N_R N_R H_{\rm SM}(\to b\bar{b})$} & $2.05 \times 10^{1}~(2.05\times 10^4)$ & $1.02 \times 10^0~(1.02\times 10^4)$ & $3.67 \times 10^{-2}~(3.67 \times 10^3)$ &  \\ 
& $5.83 \times 10^0~(5.83\times 10^3)$ & $0.31 \times 10^0~(0.31 \times 10^4)$  & $1.12 \times 10^{-2}~(1.12\times 10^3)$ & $H_{\rm SM}(\to b\bar{b}) Z(\to \nu\bar{\nu}), H_{\rm SM} \nu_\mu \bar{\nu}_\mu$  \\
& $0.47 \times 10^0~(0.47 \times 10^3)$ & $0.47 \times 10^{-1}~(0.47\times 10^3)$  & $1.81 \times 10^{-3}~(1.81 \times 10^2)$ & $t\bar{t}, Z(\to \nu\bar{\nu})Z(\to q\bar{q})$  \\ 
& $0.11 \times 10^0~(0.11 \times 10^3)$ & $0.21 \times 10^0~(0.21 \times 10^4)$  & $1.47 \times 10^{-2}~(1.47 \times 10^3)$ & \\ [0.8ex]
\bottomrule
\bottomrule
\end{tabular}
\hspace{0.2cm}
\end{adjustbox}
\end{center}
\caption{\label{tab:FS:NNX} 
The total cross sections times the branching ratio ($\sigma \times {\rm BR}$) and the expected number of signal events for the $N_R N_R$ production in association with $\gamma$, $Z(\to \ell\ell)$, $Z(\to q\bar{q})$, and $H_{\rm SM}(\to b\bar{b})$. We consider three representative center-of-mass energies of $3$, $10$ and $30$ TeV. For each process, we show four entries that correspond to the benchmark points considered in this study along with the associated background contributions. Here $\ell$ refers to either an electron or a muon.}
\end{table*}

In figure \ref{fig:XS:NR} we show the total cross sections for DM production in $\mu\mu$ collisions as function of the center-of-mass energy ($\sqrt{s_{\mu\mu}}$) for the four benchmark points defined in table \ref{tab:BSs}. Starting with mono--X processes, it is clear that the mono-$\gamma$ channel has the highest rate which varies from $\simeq 1$ pb for $\sqrt{s_{\mu\mu}} = 3$ TeV to about $80$ fb for $\sqrt{s_{\mu\mu}} = 30$ TeV in BP1\footnote{Note that for mono--$\gamma$, we have applied some generator-level cuts by requiring that $p_T^\gamma > 25$ GeV and $|\eta^\gamma| < 2.5$.}. Mono-$Z$ production has the second highest cross sections which varies between $200$ fb for $\sqrt{s_{\mu\mu}} = 3$ TeV and about $2$ fb for $\sqrt{s_{\mu\mu}} = 30$ TeV. Finally, mono-Higgs production has the lowest rates among all the mono--X processes with cross section approaching $63$ fb for $\sqrt{s_{\mu\mu}} = 3$. The rates for mono--X decrease by about a factor of 10 for BP2, by a factor of 100 for BP3 and by a factor of 10 for BP4. Notice that the decrease in the production cross sections is not due to the DM mass but also to the change in the value of $Y_{\mu N}$ since the total rates are proportional to $Y_{\mu N}^4$. An exception to this rule is in the mono-Higgs production cross section which decrease by factors of $6$--$200$ since it scales as $\lambda_3^2 Y_{\mu N}^4$. \\

The rates of the production of DM in association with two SM particles are shown in figure \ref{fig:XS:NR}. We can see that, as expected, they are suppressed as compared to the case of mono--X channels. The process with the highest is $N_R N_R + \gamma\gamma$ whose cross section is between $50$ fb and $2$ fb. This process is followed by $N_R N_R \gamma Z$ and $N_R N_R W^+ W^-$ whose cross sections are slightly smaller. An interesting process is the production of DM in association with two SM Higgs bosons whose cross sections is about $1$--$3$ fb depending on the center-of-mass energy. We note that the rates of these processes decrease as the DM mass increase, {\it i.e.} by a factor of 10 for BP2. 
\\

Finally, the production cross sections of DM in association with three SM particles are shown in figure \ref{fig:XS:NR} for BP1--BP4. It is clear that these rates are suppressed as compared to those of the DM production in association with one SM particle and two SM particles respectively. The maximum being about $1$ fb for $N_R N_R W^+ W^- \gamma$ and $N_R N_R \gamma\gamma\gamma$ at $\sqrt{s_{\mu\mu}} = 3$ TeV. We note that the dependence on $\sqrt{s_{\mu\mu}}$ of the cross sections for the production of $N_R N_R$ in association with three SM particles is not a strong as in case of other processes. Despite the smallness of these cross section, these processes may have a high sensitivity reach due to the smallness of the associated backgrounds.  

\begin{table*}[!t]
\setlength\tabcolsep{8pt}
\begin{center}
\begin{adjustbox}{max width=\textwidth}
\begin{tabular}{l c c c l}
\toprule
\toprule
& & & & \\ [-0.4ex]
 & \multicolumn{3}{c}{$\sigma \times {\rm BR}$~[fb]~({number of events})} & Dominant backgrounds \\ [2.3ex]
\toprule
\toprule
& & & & \\ [-0.4ex]
 $\sqrt{s_{\mu\mu}}$ [TeV] &  $3$ & $10$ & $30$ & \\ [2.3ex]
\toprule
\toprule
& & & & \\ [-0.2ex]
\multirow{4}{*}{$N_R N_R \gamma \gamma$} & $4.97 \times 10^{1}~(4.97\times 10^4)$ & $1.23 \times 10^1~(1.23\times 10^5)$ & $2.38 \times 10^0~(2.38\times 10^5)$ &  \\ 
& $4.93 \times 10^1~(4.93\times 10^3)$ & $1.28 \times 10^0~(1.28\times 10^4)$  & $0.25 \times 10^0~(2.53\times 10^4)$ & $\nu\bar{\nu} + \gamma\gamma, 2 \nu\bar{\nu} + \gamma\gamma$  \\
& $0.43 \times 10^0~(0.43\times 10^3)$ & $0.17 \times 10^0~(1.73 \times 10^3)$  & $0.36 \times 10^{-1}~(3.64\times 10^3)$ & $\mu\mu/VV\to H_{\rm SM}(\to\gamma\gamma)+\nu\bar{\nu}$ \\ 
& $1.00 \times 10^0~(1.00\times 10^3)$ & $1.87 \times 10^0~(1.87 \times 10^4)$  & $0.48 \times 10^0~(4.85 \times 10^4)$ & \\ [0.2ex]
\midrule
& & & &  \\ [-0.2ex]
\multirow{4}{*}{$N_R N_R \gamma Z(\to \ell\ell)$} & $1.24 \times 10^0~(1.24\times 10^3)$ & $0.49 \times 10^0~(4.98\times 10^3)$ & $1.29 \times 10^{-1}~(1.29 \times 10^4)$ &  \\ 
& $1.76 \times 10^0~(1.76\times 10^3)$ & $0.76 \times 10^0~(7.64 \times 10^3)$  & $0.20 \times 10^0~(2.02\times 10^4)$ & $\gamma Z(\to \ell\ell) + \nu\bar{\nu}$  \\
& $0.12 \times 10^{0}~(1.23 \times 10^2)$ & $9.50 \times 10^{-2}~(9.50\times 10^2)$  & $2.80 \times 10^{-2}~(2.80 \times 10^3)$ & $\gamma W (\to \ell\nu_\ell) W (\to \ell\nu_\ell)$  \\ 
& $0.18 \times 10^{0}~(1.79 \times 10^2)$ & $9.05 \times 10^{-1}~(9.05 \times 10^3)$  & $3.46 \times 10^{-1}~(3.46 \times 10^4)$ & \\ 
& & & & \\ [-0.2ex]
\midrule
& & & &  \\ [-0.2ex]
\multirow{4}{*}{$N_R N_R Z(\to \ell\ell)Z(\to \ell\ell)$} & $3.53 \times 10^{-2}~(3.53\times 10^1)$ & $2.29 \times 10^{-2}~(2.29\times 10^2)$ & $7.21 \times 10^{-3}~(7.21 \times 10^2)$ &  \\ 
& $0.98 \times 10^0~(9.80\times 10^2)$ & $0.75 \times 10^0~(7.54 \times 10^3)$  & $0.24 \times 10^0~(2.42\times 10^4)$ & $Z(\to \ell\ell)Z(\to \ell\ell)Z(\to \nu\bar{\nu})$  \\
& $3.23 \times 10^{-2}~(3.23 \times 10^1)$ & $7.87 \times 10^{-2}~(7.87\times 10^2)$  & $3.08 \times 10^{-2}~(3.08 \times 10^3)$ & $H_{\rm SM}(\to ZZ^*) Z, H_{\rm SM}(\to WW^*) Z$  \\ 
& $7.50 \times 10^{-3}~(7.50 \times 10^0)$ & $0.39 \times 10^0~(3.89 \times 10^3)$  & $0.30 \times 10^0~(3.02 \times 10^4)$ & $W(\to\ell\nu)W(\to\ell\nu)Z(\to \ell\ell)$ \\ 
& & & & \\ [-0.2ex]
\midrule
& & & &  \\ [-0.2ex]
\multirow{4}{*}{$N_R N_R V(\to q\bar{q})V(\to q\bar{q})$} & $1.05 \times 10^{1}~(1.05\times 10^4)$ & $6.57 \times 10^0~(6.57\times 10^4)$ & $2.02 \times 10^0~(2.02 \times 10^5)$ &  \\ 
& $2.76 \times 10^0~(2.76 \times 10^3)$ & $2.08 \times 10^0~(2.08 \times 10^4)$  & $0.65 \times 10^{0}~(6.57 \times 10^4)$ & $t\bar{t}, V(\to q\bar{q}) V(\to q\bar{q}) \nu\bar{\nu}$  \\
& $8.90 \times 10^{-2}~(8.90 \times 10^1)$ & $2.15 \times 10^{-1}~(2.15 \times 10^3)$  & $8.30 \times 10^{-2}~(8.30 \times 10^3)$ & $H_{\rm SM}(\to gg)H_{\rm SM}(\to gg) \nu\bar{\nu}$  \\ 
& $1.30 \times 10^{-2}~(1.30 \times 10^1)$ & $9.74 \times 10^{-1}~(9.74 \times 10^3)$  & $7.96 \times 10^{-1}~(7.96 \times 10^4)$ & $H_{\rm SM}(\to b\bar{b})H_{\rm SM}(\to b\bar{b}) \nu\bar{\nu}$\\ 
& & & & \\ [-0.2ex]
\midrule
& & & &  \\ [-0.2ex]
\multirow{4}{*}{$N_R N_R H_{\rm SM}(\to b\bar{b})H_{\rm SM}(\to b\bar{b})$} & $1.21 \times 10^{0}~(1.21\times 10^3)$ & $1.12 \times 10^0~(1.12\times 10^4)$ & $3.77 \times 10^{-1}~(3.77 \times 10^4)$ &  \\ 
& $3.95 \times 10^{-1}~(3.95\times 10^2)$ & $5.29 \times 10^{-1}~(5.29 \times 10^3)$  & $1.88 \times 10^{-1}~(1.88\times 10^4)$ & $t\bar{t}, V(\to q\bar{q}) V(\to q\bar{q}) \nu\bar{\nu}$  \\
& $1.22 \times 10^{-2}~(1.22 \times 10^1)$ & $5.32 \times 10^{-2}~(5.32\times 10^2)$  & $2.36 \times 10^{-2}~(2.36 \times 10^3)$ & $H_{\rm SM}(\to gg)H_{\rm SM}(\to gg) \nu\bar{\nu}$  \\ 
& $1.40 \times 10^{-3}~(1.40 \times 10^0)$ & $2.49 \times 10^{-1}~(2.49 \times 10^3)$  & $2.27 \times 10^{-1}~(2.27 \times 10^4)$ & $H_{\rm SM}(\to b\bar{b})H_{\rm SM}(\to b\bar{b}) \nu\bar{\nu}$ \\ [0.8ex]
\bottomrule
\bottomrule
\end{tabular}
\hspace{0.2cm}
\end{adjustbox}
\end{center}
\caption{\label{tab:FS:NNXX} 
Same as in table \ref{tab:FS:NNX} but $N_R N_R$ in association with $\gamma\gamma$, $\gamma Z(\to \ell\ell)$, $Z(\to \ell\ell) Z(\to \ell \ell)$, $V(\to q\bar{q}) V(\to q\bar{q}$ and $H_{\rm SM}(\to b\bar{b}) H_{\rm SM}(\to b\bar{b})$. Here $V$ refers to either $W$ or $Z$.}
\end{table*}

\subsection{Expected event yields and dominant backgrounds}
After discussing the total cross sections for all the possible production channels of dark matter at muon colliders, it is instructive to discuss both the total expected number of events for specific decay channels of the SM particles and the associated backgrounds. In this subsection, we focus on two categories of DM production channels ({\it i}) DM production in association with one SM particle where we consider four processes: $N_R N_R \gamma$, $N_R N_R Z(\to\ell\ell)$, $N_R N_R Z(\to q\bar{q})$ and $N_R N_R H_{\rm SM}(\to b\bar{b})$ and ({\it ii}) DM production in association with two SM particles where we consider five processes: $N_R N_R \gamma\gamma$, $N_R N_R \gamma Z(\to \ell\ell)$, $N_R N_R Z(\to \ell\ell) Z(\to \ell\ell)$, $N_R N_R V(\to q\bar{q}) V(\to q\bar{q})$ and $N_R N_R H_{\rm SM}(\to b\bar{b}) H_{\rm SM}(\to b\bar{b})$. The results are shown in tables \ref{tab:FS:NNX} and \ref{tab:FS:NNXX}. The discussion will be restricted for the following center-of-mass energies and integrated luminosities
\begin{eqnarray}
\sqrt{s_{\mu\mu}} &=& 3,~10,~{\rm and}~30~{\rm TeV} \nonumber \\ 
\int {\rm d}t {\cal L} &=& 1,~10,~{\rm and}~100~{\rm ab}^{-1},
\end{eqnarray}
where we follow ref. \cite{Han:2020uak} assuming that the luminosity has a linear scaling with the center-of-mass energy. The expected number of events for mono--X processes is calculated using the following equation
\begin{eqnarray}
{\cal N} = \sigma_{N_R N_R X} \times {\rm BR}_{X \to x_1 x_2} \times \int {\rm d}t {\cal L}.
\end{eqnarray}
For the production of DM in association with two SM particles, we have 
\begin{eqnarray}
{\cal N} = \sigma_{N_R N_R X Y} \times {\rm BR}_{X \to x_1 x_2} \times {\rm BR}_{Y  \to y_1 y_2}  \times \int {\rm d}t {\cal L}.
\end{eqnarray}

\subsubsection{$\mu^+ \mu^- \to N_R N_R + X$}

\paragraph{$N_R N_R \gamma.$}

This process leads to the final state comprising of a highly energetic photon and a large missing transverse energy ($E_{T}^{\rm miss}$). In addition one could have a few additional charged leptons, or photons that are emitted from the radiation of either the initial-state muons or the final-state photon. The dominant backgrounds for this signal process are the production of two or four neutrinos in association with a photon. The production of two neutrinos proceeds via muon-muon annihiliation -- $\mu^+ \mu^- \to Z(\to \nu\bar{\nu})\gamma$ -- and VBF -- $VV \to Z(\to \nu\bar{\nu})\gamma$ -- with cross sections varying from $2.98$ pb for $\sqrt{s_{\mu\mu}} = 3$ TeV to $3.27$ pb for  $\sqrt{s_{\mu\mu}} = 30$ TeV. The production of four neutrinos in association with hard photon has an extremely cross section with the maximum being $1.5$ fb for $\sqrt{s_{\mu\mu}} = 30$ TeV. It is worth noting from table \ref{tab:FS:NNX} that the signal significance can easily $5$\footnote{The signal significance is defined as ${\cal S}/\sqrt{{\cal B}}$ with ${\cal S}$ is the number of signal events and ${\cal B}$ is the number of background events.}. For the other benchmark points, a more detailed selection is required to reach a signal significance of $5$ if one can achieve an acceptance times efficiency ($A\times \epsilon$) of about $15\%$ for the signal in the signal region while the background is having $A\times \epsilon$ of about $\mathcal{O}(10^{-3})$. \\

 \begin{figure*}[!t]
\centering
\includegraphics[width=0.49\linewidth]{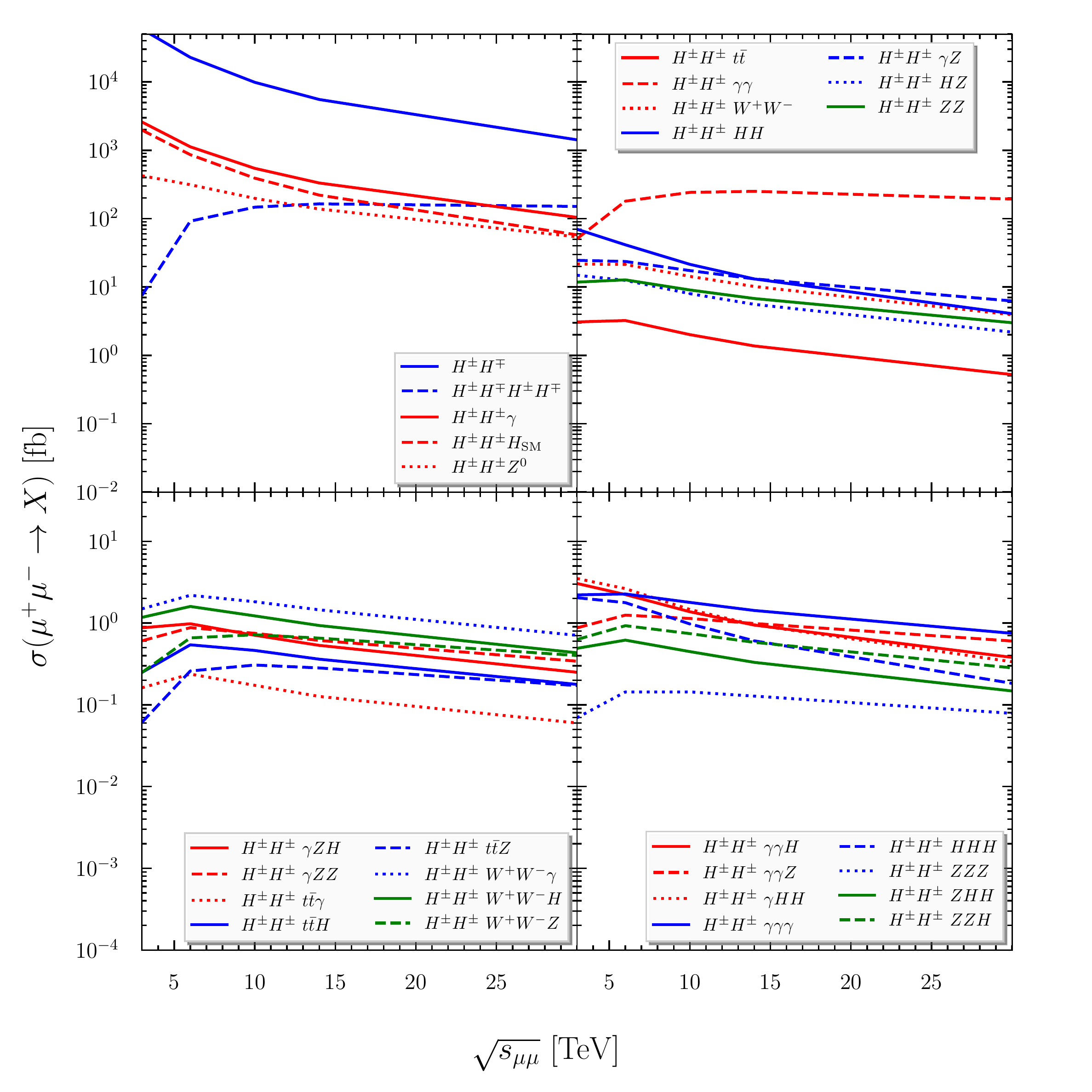}
\hfill
\includegraphics[width=0.49\linewidth]{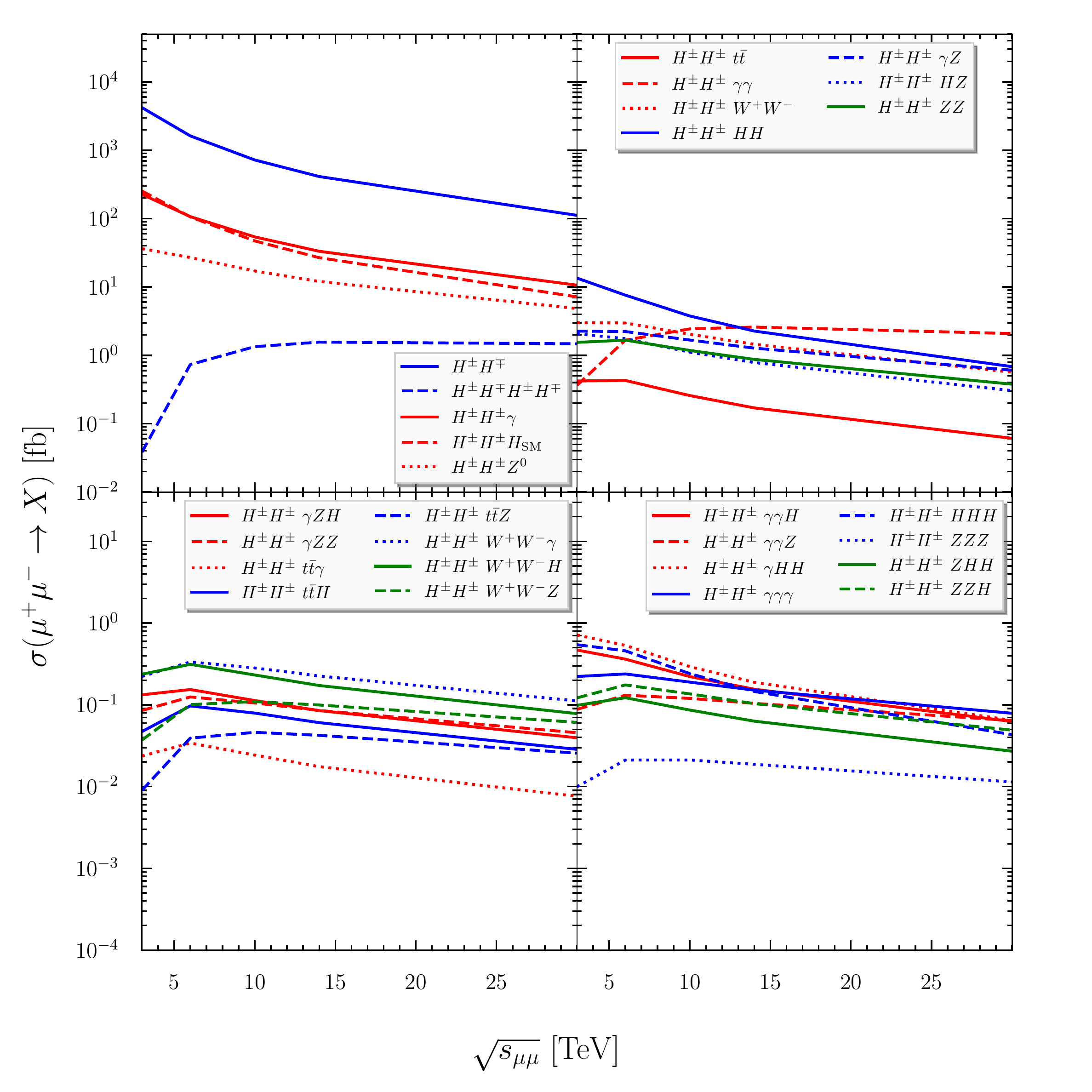}
\vfill
\includegraphics[width=0.49\linewidth]{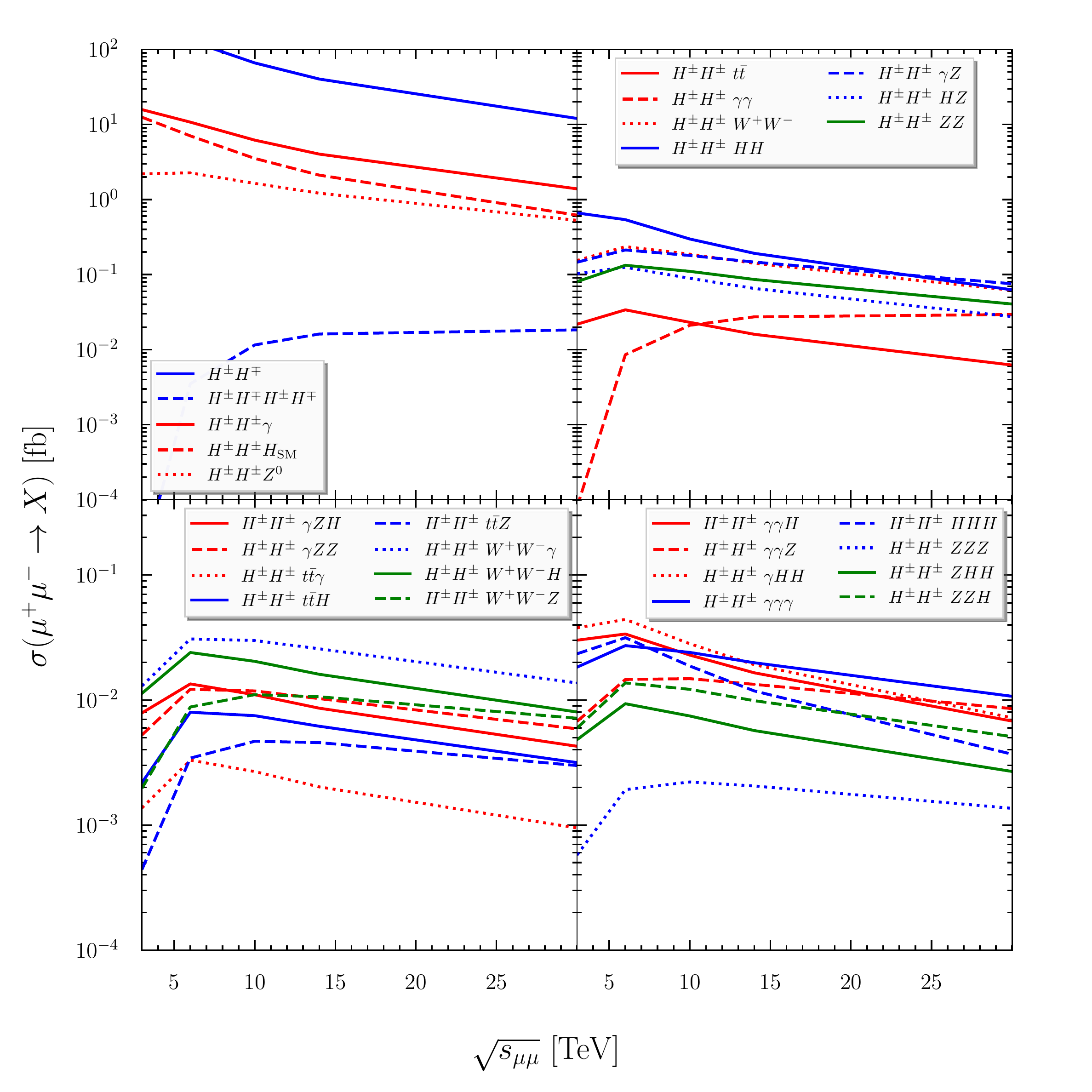}
\hfill
\includegraphics[width=0.49\linewidth]{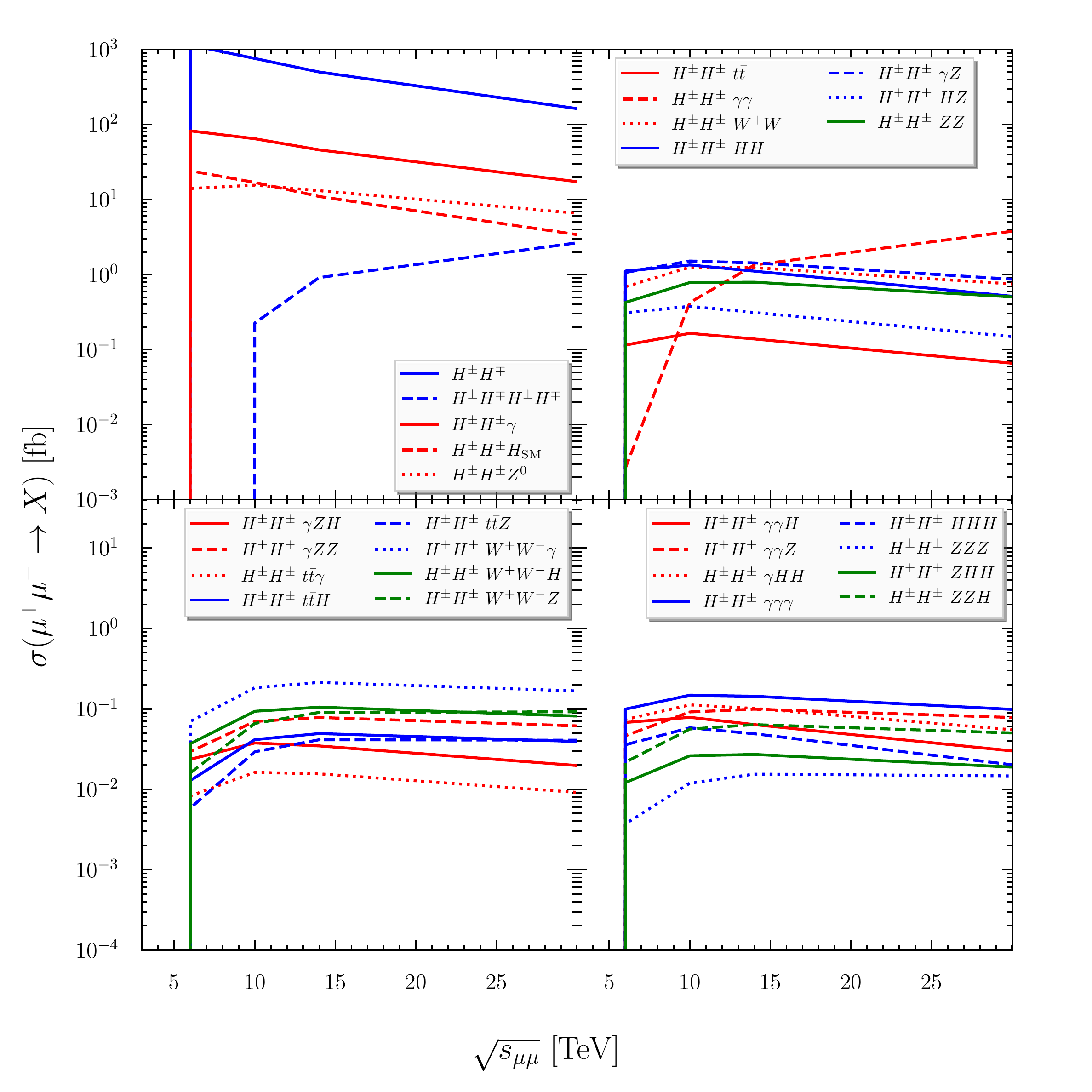}
\caption{Production cross section of $H^\pm H^\mp + X$ as a function of the center-of-mass energy ($\sqrt{s_{\mu\mu}}$) for the benchmark points BP1 (left upper panel), BP2 (right upper panel), BP3 (left lower panel) and BP4 (left lower panel). For each pane, we show the production cross section for $H^\pm H^\mp$ plus one SM particle, plus two SM particles and in association with three SM particles.}
\label{fig:XS:H}
\end{figure*}  

\paragraph{$N_R N_R Z(\to \ell \ell).$}

This process leads to a very clean final state containing two same-flavour opposite-sign (SFOS) charged leptons from the decay of the $Z$-boson in association with large missing energy. The dominant backgrounds are found to be the production of two $Z$-bosons with one decaying two charged leptons and the other decaying invisibly. We note that there is another background originated from the production of two $W$-bosons both decaying leptonically which can significantly be reduced using the requirement of two SFOS leptons whose invariant mass is close to the $Z$-boson mass. The cross sections for the $ZZ$ production varies from $0.4$ fb to $26$ fb for the muon-muon annihilation (decreases while the center-of-mass energy increases) and from $56$ fb to about $430$ fb (increases with the center-of-mass energy). On the other hand, the cross section for $WW$ production is larger and varies between $8.5$ fb and $466$ fb in the muon annihilation channel and between $150$ fb and $858$ fb in the VBF channels. The expected number of events for this signal process is about ${\cal O}(10^3$--$10^4)$. Given the differences in the topology of the signal and backgrounds, it is easy to achieve a significance of $5\sigma$ by suitable event selection. \\

\paragraph{$N_R N_R Z(\to q\bar{q})$ and $N_R N_R H_{\rm SM}(\to b\bar{b}).$}

This category of channels involves two hadronic jets in association with missing energy. For the case of the $Z$--boson, the main decay channel is into $q\bar{q}; q=u,d,s,c,b$ with ${\rm BR}(Z\to q\bar{q})=69.911\%$ \cite{ParticleDataGroup:2022pth}. For the SM Higgs boson, the main decay is into $b\bar{b}$ with ${\rm BR}(H_{\rm SM} \to b\bar{b}) = 57\%$. The dominant backgrounds to these signal processes come from $q\bar{q}$ production in association with two neutrinos, SM Higgs boson production, $t\bar{t}$ production with one top quark decaying leptonically and the other top decaying hadronically and $WW$ production where one $W$-boson decays leptonically and the other decaying hadronically. In the last two backgrounds, one needs that the charged lepton escapes the detection volume. Since the both the hadronically decaying $Z$-- and Higgs-bosons are accompanied with very large missing energy, their decays are not always resolved as two well separated two jets but rather a fat jet with specific characteristics. We expect for these channels a decent statistics and the backgrounds are under control with suitable selection.

\subsubsection{$\mu^+ \mu^- \to N_R N_R + XY$}

\paragraph{$N_R N_R \gamma\gamma$.}

In this model, there is a possibility to produce DM pairs in association with two hard photons. The expected final state would consist of two hard photons in addition to large missing energy. Contrary to mono--$\gamma$ channel, this process does not have large backgrounds where the main backgrounds are the production of two photons in association with two neutrinos: non-resonant and resonant (through the decay of the SM Higgs boson). The resonant backgrounds can be easily suppressed via suitable requirements on the invariant mass of the diphoton system, {\it i.e.} removing photons that are within the SM Higgs mass window. For all the benchmark points we expect a decent statistics for the signal events, {\it i.e.} of about ${\cal O}(10^3$--$10^5)$. This would imply that this process would one of the golden modes to probe DM at muon colliders which will be studied in details in a future work. \\

\paragraph{$N_R N_R \gamma Z(\to \ell\ell)$.}

This is also one of the unique processes to probe DM at muon colliders. The final state consists of one hard photons, two charged leptons and large missing transverse energy. The associated background is manageable since it consists of the production of one photon and one or two gauge bosons. The expected number of signal events is quite large as well, {\it i.e.} of about ${\cal O}(10^2$--$10^4)$.  \\

\paragraph{$N_R N_R Z(\to \ell\ell) Z(\to \ell\ell)$.}
This one of the most cleanest final states that can be used to probe DM at muon colliders. The signature consists of four charged leptons in association with missing energy. The corresponding is even smaller than for the other signal processes. We note that enough statistics can be only achieved at $\sqrt{s_{\mu\mu}} = 30$ TeV where we expect about ${\cal O}(10^2$--$10^4$) events. The major backgrounds arise from the production of three gauge bosons or from the production of the SM Higgs boson decaying into $VV^*, V=W,Z$ in association with one or two gauge bosons. \\

\paragraph{$N_R N_R V(\to q\bar{q}) V(\to q\bar{q})$ and $N_R N_R H_{\rm SM}(\to b\bar{b}) H_{\rm SM}(\to b\bar{b})$.}

The production of two gauge bosons or two SM Higgs bosons in association with DM pairs lead to purely hadronic final states (either four resolved jets or two fat jets) in association with large missing energy. The dominant backgrounds for these signal processes consist of the production of two SM neutrinos in association with two gauge bosons, two SM Higgs bosons, or one Higgs boson and one gauge boson decaying hadronically. This process will be studied in great detail in a future work.

\section{Production of charged scalars at muon colliders}
\label{sec:prod:S}

In this section, we discuss the production of charged scalar pairs at muon colliders. Similarly to the production of DM, charged scalars can be produced either in association with one SM particle, with two SM particles or three SM particles. In addition, we could have the production of charged scalar pairs with non SM particles ($H^\pm H^\pm$) or the production of four charged scalars. An interesting feature about the production of charged scalars is that the appearance of at least two leptons in association with missing energy in addition to the decay products of the SM particles. For example, the production of charged scalar pairs in association with a SM Higgs boson would lead to two hard charged leptons, missing energy and two $b$-tagged jets (or one fat jet). On the other hand, the charged scalar production receives contributions from VBF thanks to their couplings to $\gamma/Z$. The results of the production cross sections for the different processes involving charged scalars as a function of the center-of-mass energy are shown in figure \ref{fig:XS:H}. Below, we list the possible production channels for the charged scalars: \\

\paragraph{$\mu\mu \to H^\pm H^\mp/H^\pm H^\mp H^\pm H^\mp$.} These processes lead to signatures of either two charged lepton and MET or four charged leptons and MET. Charged scalar pair production proceeds through either $s$--channel diagrams with the exchange $\gamma/Z$--bosons or $t$--channel diagram with the exchange of the Majorana DM. The cross section for charged scalar pair production ranges from about $10^4$ fb to about $10^1$ fb. It is worth noting that for the benchmark point BP3 has the smallest cross section due to the tiny mass splitting of about $2$ GeV between the charged scalar and the DM candidate. In all the case, the number of events for this process is quite large. The cross section for charged scalar pair production has a $1/\sqrt{s_{\mu\mu}}$ scaling. The cross section for the production of four charged scalars is smaller as expected due to phase suppression. It is however quite decent as can be seen in fig. \ref{fig:XS:H} and ranges from $10^{-2}$ fb to $10^2$ depending on the benchmark scenarios. The most notable signatures are $4$ muons plus MET (BP1, BP2, BP3) and $2$ muons and $2$ tau leptons (BP4). These two channels will be studied in great detail in a future work \cite{Jueid:2023xyz}. \\

\paragraph{$\mu\mu \to H^\pm H^\mp + X$.} In this case, we have three production channels: $H^+ H^- \gamma$, $H^+ H^- Z$ and $H^+ H^- H_{\rm SM}$.  There are three contributions to $H^+ H^- \gamma$: $s$--channel contributions through $\gamma/Z$ with the photon being emitted from the $H^+ H^-$ vertex and $t$--channel contribution through the exchange of $N_R$. The final-state signature for this process consists of two charged leptons in association with one hard photon (the kinematics is quite different from $N_R N_R \gamma Z$ production). We can see in fig. \ref{fig:XS:H} that the cross section ranges from $10^1$ fb to $10^3$ fb depending on the center-of-mass energy and the benchmark point. Secondly, we can have the production of charged scalar pairs in association with one $Z$-boson which would lead to very rich signatures: $2\ell$ + MET, $2\ell$ + 2 jets + MET or $4\ell$ + MET. The cross sections for these processes are shown in fig. \ref{fig:XS:H} where it is clear that the rates are quite important from $10^0$ to $10^2$ fb. Finally, the charged scalar pairs can be produced in association with a SM Higgs boson.  The rates for this interesting channel are also quite important and range between $10^0$ fb and $10^2$ fb. \\

\paragraph{$\mu\mu \to H^\pm H^\mp + XY$.} For this category we have seven production channels: $H^+ H^- \gamma\gamma$, $H^+ H^- \gamma Z$, $H^+ H^- Z Z$, $H^+ H^- W^+ W^-$, $H^+ H^- H_{\rm SM} H_{\rm SM}$, $H^+ H^- H_{\rm SM} Z$ and $H^+ H^- t\bar{t}$. The rates for this channels are quite smaller but still at the noticeable level, {\it i.e.} from $10^{-2}$ fb to $10^2$ fb depending on the center-of-mass energy and the benchmark point. It is worth noting that the production of charged scalar pairs in association with two SM particles leads to even much more richer signatures with very small backgrounds, {\it it.e.} $6$ leptons plus MET, $4$ leptons plus 4 jets plus MET and so on. \\

\paragraph{$\mu\mu \to H^\pm H^\mp + XYZ$.} This is the most complicated category of processes where we can have 16 processes with many more final-state signatures. The rates for these processes are much more smaller with the maximum being about $3$ fb for $H^+ H^- \gamma H_{\rm SM} H_{\rm SM}$ and $H^+ H^- \gamma \gamma H_{\rm SM}$ at $\sqrt{s_{\mu\mu}} = 3$ TeV. \\

\begin{figure}[!t]
    \centering
    \includegraphics[width=0.95\linewidth]{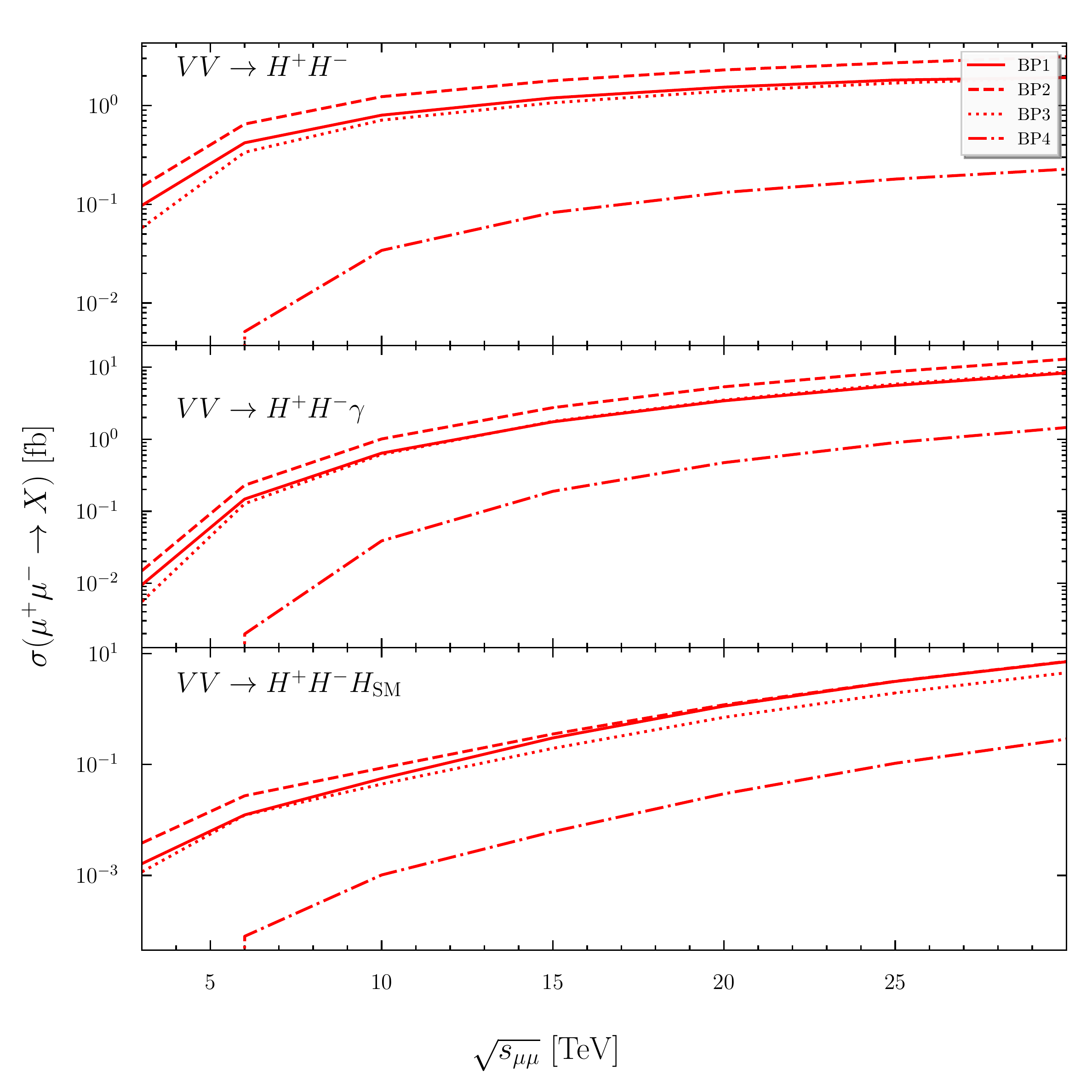}
    \caption{The production cross sections for $H^+ H^-$, $H^+ H^- \gamma$ and $H^+ H^- H_{\rm SM}$ through VBF as a function of the center-of-mass energy ($\sqrt{s_{\mu\mu}}$) for BP1 (solid), BP2 (dashed), BP3 (dotted) and BP4 (dash-dotted).}
    \label{fig:XS:VBF:H}
\end{figure}

We close this section by a brief discussion of the contribution of VBF to the production of charged scalars in this model. As mentioned earlier the charged scalar couples to the photon and the $Z$-boson and therefore may receive pure gauge VBF contributions to the total production cross section. In this model, we can have the production of charged scalars through $\gamma\gamma H^+ H^-$, $\gamma Z H^+ H^-$, $Z Z H^+ H^-$, $ZZ\to H_{\rm SM} \to H^+ H^-$ and $W^+ W^- \to \gamma/Z \to H^+ H^-$ vertices. We take examples of production of $H^+ H^-$, $H^+ H^- \gamma$ and $H^+ H^- H_{\rm SM}$ and show the corresponding results for the four benchmark points in fig. \ref{fig:XS:VBF:H}. We can see that the cross sections increase with center-of-mass energy but do not go above $2$ fb for $H^+ H^-$ in BP2. Therefore, the muon annihilation channels are the most important in our model thanks to the $Y_{\mu N}^4$  dependence of the cross section. 

\section{Conclusions}
\label{sec:conclusions}

In this work we have studied the production of DM and charged scalars at high energy muon colliders within the minimal lepton portal DM model.  The model consists of extending the SM with two $SU(2)_L$ gauge singlets: a charged singlet scalar and a neutral right-handed  fermion (or equivalently a Majorana fermion). We first discussed in details the phenomenology of the model at the LHC and the corresponding constraints from direct detection, relic density measurement, and lepton flavour violating decays of charged leptons and the SM Higgs boson. Then we have selected a few benchmark points that define some phenomenologically viable scenarios and which can be tested at future muon colliders. For these benchmark points, we have calculated the cross sections for the production of DM in association with SM particles and of charged scalars of the models in association with SM particles as a function of the center-of-mass energy. For DM production in association with SM particles, we have studied the total rates of 26 possible channels for the  benchmark points considered in this study. Furthermore, we studied the total number of events and the associated backgrounds for 9 prominent channels and found that they are very important for the  discovery DM at muon colliders for masses up to $\sim 1$ TeV.  We furthermore analysed the production of charged scalar production in association with SM particles (about 28 channels). The potential discovery for DM through charged scalar production at muon colliders is also as interesting as for direct production of DM. Further investigations of this model at muon colliders are ongoing where a full signal-to-background optimisation will be carried out for a number of  selected channels.

\section*{Acknowledgements}
The work of AJ is supported by the Institute for Basic Science (IBS) under the project code, IBS-R018-D1.

\bibliographystyle{utphys}
\bibliography{main.bib}

\end{document}